\documentclass[aip,amsmath,amssymb,reprint]{revtex4-1}
\usepackage{graphicx}% Include figure files
\usepackage{dcolumn}% Align table columns on decimal point
\usepackage{bm}% bold math
\usepackage{subfigure}
\usepackage{amsmath}
\usepackage{cases}
\usepackage[utf8]{inputenc}
\usepackage[T1]{fontenc}
\usepackage{mathptmx}
\usepackage{appendix}
\usepackage[colorlinks,linkcolor=blue,anchorcolor=blue,citecolor=blue]{hyperref}
\usepackage[nameinlink,noabbrev]{cleveref}
\usepackage{txfonts}
\usepackage{url}

\begin{document}

\preprint{AIP/123-QED}

\title{The tipping times in an Arctic sea ice system under influence of extreme events}
% Force line breaks with \\

\author{Fang Yang}
\affiliation{
	Center for Mathematical Science, School of Mathematics and Statistics,\\ Huazhong University of Science and Technology, Wuhan, 430074, China.%\\This line break forced with \textbackslash\textbackslash
}%
\author{Yayun Zheng}%
\affiliation{
	Center for Mathematical Science, School of Mathematics and Statistics,\\ Huazhong University of Science and Technology, Wuhan, 430074, China.%\\This line break forced with \textbackslash\textbackslash
}%
\affiliation{ Wuhan National Laboratory for Optoelectronics, Huazhong University of \\ Science and Technology, Wuhan, 430074,China.
}
\email{Corresponding author: yayunzh55@hust.edu.cn}
\author{Jinqiao Duan}
\affiliation{
	Department of Applied Mathematics, Illinois Institute of Technology, \\Chicago, IL 60616, USA.%\\This line break forced% with \\
}

\author{Ling Fu}
\affiliation{ Wuhan National Laboratory for Optoelectronics, Huazhong University of \\ Science and Technology, Wuhan, 430074,China.
}
\author{Stephen Wiggins}
\affiliation{
	School of Mathematics, University of Bristol, Fry Building, Woodland Road, BRISTOL BS8 1UG, United Kingdom.
	%\\This line break forced% with \\
}

\date{\today}% It is always \today, today,

\begin{abstract}

In light of the rapid recent retreat of Arctic sea ice, the extreme weather events triggering the variability in Arctic ice cover has  drawn increasing attention. A non-Gaussian $\alpha$-stable  L\'evy process is  thought to be an appropriate model to describe such extreme event. The maximal likely trajectory, based on the nonlocal Fokker-Planck equation, is applied to a  nonautonomous  Arctic sea ice system under $\alpha$-stable  L\'evy noise. Two types of  tipping times, the early-warning tipping time and the disaster-happening tipping time, are used to predict the critical time for the maximal likely transition from a perennially ice-covered state to a seasonally ice-free one, and from a seasonally ice-free state to a perennially ice-free one, respectively.  We find that the increased  intensity of extreme events results in shorter warning time for sea ice melting, and that an enhanced greenhouse effect will intensify this influence, making the arrival of warning time significantly earlier. Meanwhile, for the enhanced greenhouse effect, we discover that increased intensity and frequency of extreme events will advance the disaster-happening tipping time, in which an ice-free state is maintained throughout the year in the Arctic Ocean. Finally, we identify values of L\'evy index $\alpha$ and noise intensity $\epsilon$ in $\alpha \epsilon$-space that can trigger a transition between the Arctic sea ice state. These results  provide an effective theoretical framework for studying Arctic sea ice variations  under the influence of extreme events.

\end{abstract}

\maketitle

\begin{quotation}
	One of the most dramatic indicators of Arctic warming has been the decline in the sea ice cover. To gain insight into whether Arctic sea ice under extreme weather events will seasonally or completely disappear in the future, we consider an Arctic sea ice model driven by a non-Gaussian $\alpha$-stable L\'evy process. In this paper, we use the maximal likely trajectory obtained from the nonlocal Fokker-Planck equation to characterize the most likely evolution process of  Arctic sea ice under  L\'evy noise. We introduce the early-warning tipping time (the time for breaking the ice-covered state) and the disaster-happening tipping time (the time for beginning the perennially ice-free state) to predict the variability of Arctic sea ice.  Finally, we find values of L\'evy index $\alpha$ and L\'evy noise intensity $\epsilon$ that can activate  a transition from one stable state to the other in this Arctic sea ice system.

\end{quotation}

\section{\label{sec:level1}Introduction}

Arctic sea ice variations are important indicators of climate changes\cite{Min2008, Notz2012}. Satellite observations have revealed a substantial decline in September Arctic sea ice extent since the late 1970s \cite{Schweiger2019}.   Several studies \cite{Budyko1969, Sellers1969, North1981, Crowley2000, Zhang2000} have shown that the generation and melting of sea ice are affected by energy flux involving seasonal variations in solar radiation, thermodynamics, and heat transport in the atmosphere and ocean. Due to the complexity of the overall system, it is necessary to use mathematical models to determine how these physical processes interact in order to provide explanations for both observed and possible future behaviors of the time evolution of Arctic sea ice.

In the context of energy flux balance models, Budyko \cite{Budyko1969} and Sellers \cite{Sellers1969} have recognized the advantages of simple deterministic theories of climate that provide a clear assessment of stability and feedbacks \cite{Moon2017}. Afterwards, a deterministic single-column energy flux balance model  has been proposed for the Arctic Ocean by Eisenman and Wettlaufer \cite{Eisenman2009}. Subsequently, this model has been improved by including additional physical mechanisms, such as the influence of clouds $\cite{Abbot2011,Hill2016}$, in the time-dependent terms of the equation.

 There are two reasons to include random variables and/or processes in a deterministic model: (i) to provide a statistical model for differences between the output of the deterministic model and observational data, such as in time series models (for example, ARMA models) of observational data \cite{Gao2016}, and (ii) to circumvent the problem of accounting for environmental influences that are too complex to include in the deterministic equations \cite{Van1976}.

 External disturbances that fluctuate rapidly relative to slowly responding state variables are often treated as stochastic processes with very short correlation times. For example, Hasselman \cite{Hasselman1976} showed that the evolution of slowly varying ``climate variables'' subject to short-time, fluctuating ``weather variables'' can be approximated by diffusion processes. In these models, climate variability, driven by short time-scale fluctuations, may be large or small depending on the degree of stabilizing feedback. A common feature of observations is that ice extent exhibits Gaussian noise structure on annual to biannual time scales \cite{Agarwal2018}. Consequently, a stochastic Arctic sea ice model with Brownian motion has been considered. This stochastic model can be used to quantify how white noise impacts the potential transition between the  ice-covered state and the ice-free one\cite{Moon2017, Chen2019}. Recently, increasing attention has been given to extreme events that trigger the variations and evolutions of  Arctic sea ice. Extreme weather events, such as heatwaves, droughts, floods, hurricanes, blizzards and other events, that occur rarely and unpredictably, can have a significant impact on climate changes and human survival. Meanwhile, it is reported that  extreme events of the type that occur in weather and climate fluctuations are more appropriately modeled as realizations of power law, or heavy tailed, distributions \cite{Farazmand2017, Selmi2016}.  These  characteristics of extreme events, described above, have strong non-Gaussianity, which cannot be described by general Gaussian noise.  A L\'evy process  is  thought to be an appropriate model for  such non-Gaussian fluctuations, with properties such as  intermittent jumps and heavy tail. Furthermore,  researchers have indicted that the presence of   $\alpha$-stable  L\'evy noise  could imply that the  underlying mechanisms for  abrupt climatic changes are single extreme events \cite{Ditlevsen1999, Zheng2020}.   We  therefore consider an Arctic sea ice model under influence of $\alpha$-stable non-Gaussian L\'evy noise in the following study.

In view of the rapid decrease in summer sea ice extent in the Arctic Ocean during the past decade \cite{Livina2013}, a potential tipping point in summer sea ice has been considered in recent studies of dynamical systems. The term tipping point commonly refers to a critical threshold at which a tiny perturbation can qualitatively alter the state or development of a system \cite{Leton2008}. Tipping points  associated with bifurcations or induced by noise are studied in a simple global energy balance model\cite{Ashwin2012, Sutera1981,Lucarini2017,Lucarini2019}. 
 
 In this paper, we use the maximal likely trajectory to determine  tipping times for the most probable transitions from a perennially ice-covered state to a seasonally ice-free one, and from a seasonally ice-free state to a perennially ice-free one, respectively. We expect that the most probable tipping time serves as a valid indicator to predict when Arctic sea ice begins to melt, and to estimate the transition time to the most devastating state corresponding to when the Arctic sea ice melts completely.

The structure of this paper is as follows. We present our methods, and introduce an Arctic sea ice model driven by $\alpha$-stable L\'evy process in Section \ref{sec:level2}.  Then we conduct numerical experiments      to investigate the impact  of non-Gaussianity and the greenhouse effect on the tipping times for  transitions in this stochastic Arctic sea ice model in Section \ref{sec:level3}. Concluding remarks appear in Section \ref{sec:level4}. In Appendix A we provide a brief description of the symmetric $\alpha$-stable L\'evy process. Appendix B contains definitions and nominal values of parameters used in our calculations.

\section{\label{sec:level2}  METHODS and MODEL} 
 
 In this section we define the maximal likely trajectory for a stochastic dynamical system driven by $\alpha$-stable L\'evy noise. The definition is based on the solution of its associated nonlocal Fokker-Planck equation which is presented below. We then introduce a model for stochastic Arctic sea ice subject to extreme weather events which, in turn, are modeled by $\alpha$-stable L\'evy noise.

%%%%%%%%%%%%%%%%%%%%%%%%%%%%%%%%%%%%%%%%%%%
\bigskip
\textbf{A. Nonlocal Fokker-Planck equation for the probability density}

\bigskip
 Consider the following scalar nonautonomous  stochastic differential equation (SDE):
\begin{equation}\label{e1}
{\rm d} X(t) = f(X(t), t) {\rm d} t + \epsilon {\rm d} L^{\alpha}_{t}, \quad X(0)=x_0 \in \mathbb{R},
\end{equation}
%where $X(t)$ is a $\mathbb{R}$-valued stochastic process. The nonautonomous drift term $f: \mathbb{R}\times \mathbb{R} \rightarrow \mathbb{R}$  satisfies some Lipschitz or H${\rm \ddot{o}}$lder continuous conditions to ensure the existence and uniqueness of the solution of the  SDE $({\ref{e1}})$. $L^{\alpha}_{t}$ is a symmetric $\alpha$-stable L\'evy process with  $\alpha \in (0, 2)$ defined on the  probability space $(\Omega, \mathcal{F}, P)$ and it is a pure jump process. The detailed introduction for  $\alpha$-stable L\'evy process is given by Appendix A. Specially, when $\alpha=2$, the symmetric $\alpha$-stable process is simply a Brownian motion, which is a Gaussian process.
where $X(t)$ is an $\mathbb{R}$-valued stochastic process. Here $L^{\alpha}_{t}$ is a symmetric $\alpha$-stable L\'evy process, with  $\alpha \in (0, 2]$, defined on the  probability space $(\Omega, \mathcal{F}, P)$ (See Appendix A).  The positive  noise intensity is $\epsilon $. The nonautonomous drift term $f: \mathbb{R}\times \mathbb{R} \rightarrow \mathbb{R}$  satisfies a Lipschitz    condition to ensure the existence and uniqueness of the solution of the  SDE $({\ref{e1}})$. A symmetric $2$-stable process is a Brownian motion, which is a Gaussian process. When L\'evy index $\alpha \in (0, 2)$, the $\alpha$-stable L\'evy process is a   jump process. A brief introduction to the $\alpha$-stable L\'evy process is given in Appendix A.

For $x_0 \in \mathbb{R}$, we suppose that the nonautonomous SDE $(\ref{e1})$ has a unique strong solution, and the probability density for this solution exists and is strictly positive. The probability density function $p(x,t)  \triangleq p(x,t| x_0,0)$ of the solution process $X(t)$ driven by a non-Gaussian $\alpha$-stable L\'evy process satisfies the following nonlocal Fokker-Planck equation \cite{Duan2015, Sun2012}:

\begin{equation}
\begin{split}
\dfrac{{\rm d}}{{\rm d}t} p(x,t) =&- \dfrac{\rm \partial }{ \partial x}\left ( f(x,t)p(x,t)\right)\\
&+\epsilon ^{\alpha} \int_{\mathbb{R}  \setminus\lbrace0\rbrace } \left( p(x+y,t)-p(x,t)-I_{\{y<1\}} \partial_{x}p(x,t) \right) \nu_{\alpha}({\rm d}y),\\
\end{split}
\label{e2}
\end{equation}
with initial condition $p(x,0)=\delta{(x-x_0)}$.

When $\alpha=2$, the SDE (\ref{e1}) is driven by Brownian motion. In this case the density function $p(x,t)$ is the solution of the Fokker-Planck equation \cite{Duan2015}:
 
\begin{equation}\label{ee3}
\begin{split}
\dfrac{{\rm d}}{{\rm d}t} p(x,t) =&- \dfrac{\rm \partial }{ \partial x}\left ( f(x,t)p(x,t)\right)
+\frac{\epsilon ^{2}}{2}\dfrac{\rm \partial ^2}{ \partial x^2}p(x,t).\\
\end{split}
\end{equation}
We use the same initial condition for equation (\ref{ee3}) that we used for equation (\ref{e2}), namely $p(x,0)=\delta{(x-x_0)}$.

In the present paper, we use the ``punched-hole" trapezoidal numerical algorithm of Gao \emph{et al.}\cite{Gao2016b} to find the  solution of the nonlocal Fokker-Planck equation (\ref{e2}).
%%%%%%%%%%%%%%%%%%%%%%%%%%%%%%%%%%%%%%%%%%

\bigskip
\textbf{B. The maximal likely trajectory }
\bigskip

When it comes to trajectory for a nonautonomous SDE, one likely option is to plot sample solution orbits from an initial state, mimicking deterministic phase portraits. However, each sample solution trajectory is an ``outcome'' of a trajectory, which could hardly offer useful information for understanding dynamics. So we consider the maximal likely trajectory, which is determined by the maximizes of the probability density function $p(x,t)$, at every time $t$.

Given $X(0)=x_0$, the maximal likely state $\cite{Zeitouni1987,Zeitouni1988}$ $x_m(t)$ of the stochastic system (\ref{e1}) at each time $t\in [0, T_f]$ is defined by 

\begin{equation} \label{e3}
x_m(t) = \underset{x\in \mathbb{R}}{{\rm argmax}} ~p(x,t|x_0,0).
\end{equation}
Here the maximizer $x_m(t)$ for $p(x,t)$ indicates the most probable location of these orbits at time $t$.

Thus, for the point $t_i$, based on equations $(\ref{e2})$ and $(\ref{e3})$, we get the maximal likely state $x_m(t_i) $ by computing the maximum of $p(x, t_i)$. We connect this series of $\{ (x_{m}(t_i),~ i=1,2,\ldots \}$ to get the maximal likely trajectory. The distances between the mesh points $\{t_i,~ i=1,2,\ldots\}$ need to be sufficiently small to get a good approximation of the maximal likely trajectory. Note that the maximal likely trajectory $x_m(t)$ is not a solution of nonautonomous SDE (\ref{e1}).

%%%%%%%%%%%%%%%%%%%%%%%%%%%%%%%%%%%%%%%%%%
\bigskip
\textbf {C. Arctic sea ice model}
\bigskip

We consider a model for  Arctic sea ice established by  Eisenman and Wettlaufer \cite{Eisenman2009}. This model is based on energy balance model where the energy per unit surface area, $E$ (with units $Wm^{-2}yr$).  When $E<0$, the layer is interpreted to consist entirely of ice. Conversely, when $E \geq 0$, the layer is interpreted to consist entirely of water, that is, the layer is in its ice-free state.  In this model $E$ satisfies the following nonautonomous differential equation \cite{Eisenman2009}:

\begin{equation}
\small
\frac{dE}{dt} =\left(1-\alpha (E)\right)F_{S}(t)-F_{0}(t)-F_{T}(t)T(t,E)+\Delta F_{0}+F_{B}+\nu_{0} \mathit{R} (-E),
\label{e4}
\end{equation}
  where

  \begin{equation*}
  \small
  \alpha (E)=\dfrac{\alpha_{ml}+\alpha_{i}}{2}+\frac{\alpha_{ml}-\alpha_{i}}{2} {\rm tanh}\left(\frac{E}{L_{i}H_{\alpha}}\right),
  \label{Eq2}
  \end{equation*}
and

  \begin{equation*}
  \small
  T(t,E)=\begin{cases}
  -\mathit{R}\left[ \dfrac{(1-\alpha_i)F_{s}(t)-F_{0}(t)+\Delta F_{0}}{k_{i}L_{i}/E-F_{T}(t)}\right] &~~ E <0,\\
  \dfrac{E}{c_{ml}H_{ml}} &~~  E\geq 0.\\
  \end{cases}
  \label{Eq3}
  \end{equation*}
  Here $\alpha(E)$ is the surface albedo, which is a central control element involved in transition dynamics.  The fraction $(1-\alpha(E))F_{S}(t)$ is the amount of absorbed short-wave radiation by the ice albedo $\alpha_{i}$ and the ocean albedo $\alpha_{ml}$. The deterministic term $\alpha(E)$ describes the energy flux balance at the atmosphere ice (ocean) interface where we calculate the surface temperature $T(t, E)$. Specifically, the state variable $E$ has the physical interpretation that the energy is stored in sea ice as latent heat when the ocean is in the ice-covered state (i.e. $E<0$) or in the ocean mixed layer as sensible heat when the ocean is in the ice-free state (i.e. $E\geq 0$). The term $\Delta F_{0}$ represents the reduction in outgoing long-wave radiation due to increased greenhouse gas forcing levels. Incident surface short-wave radiation $F_S(t)$ and basal heat flux $F_B$ are specified at central Arctic values \cite{Maykut1971}. The final term $\nu_0 \mathit{R}(-E)$ in equation $(\ref{e4})$ is the fraction of sea ice pushed by wind out of the Arctic each year. $\mathit{R}(-E)$ ensures this term is zero when there is no sea ice, where

  \begin{equation*}
  \mathit{R}(x)=\begin{cases}
  0 ,&~~ x <0,\\
  x,&~~  x\geq 0.\\
  \end{cases}
  \label{Eq4}
  \end{equation*}

The seasonally varying parameters $F_{0}(t)$ and $F_{T}(t)$, which are used to determine the surface energy flux, have values computed by using a atmospheric model \cite{Eisenman2009}. Nominal values and additional details about the model parameters are described in Appendix B (Table (\ref{tab:table1})).

\begin{figure}
	\centering
	\subfigure[]
	{
		\label{fig1:subfig:a}%
		\includegraphics[width=0.4\textwidth]{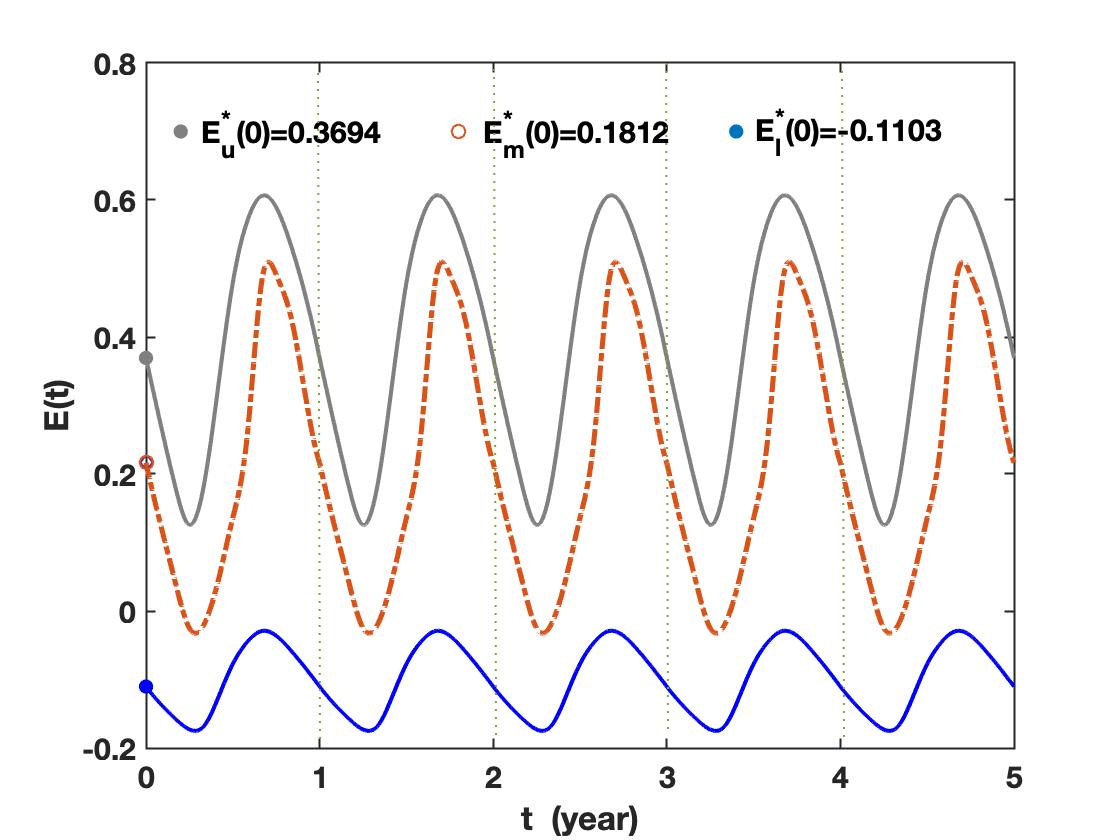}
	}
	\subfigure[]
	{	
		\label{fig1:subfig:b}%
		\includegraphics[width=0.4\textwidth]{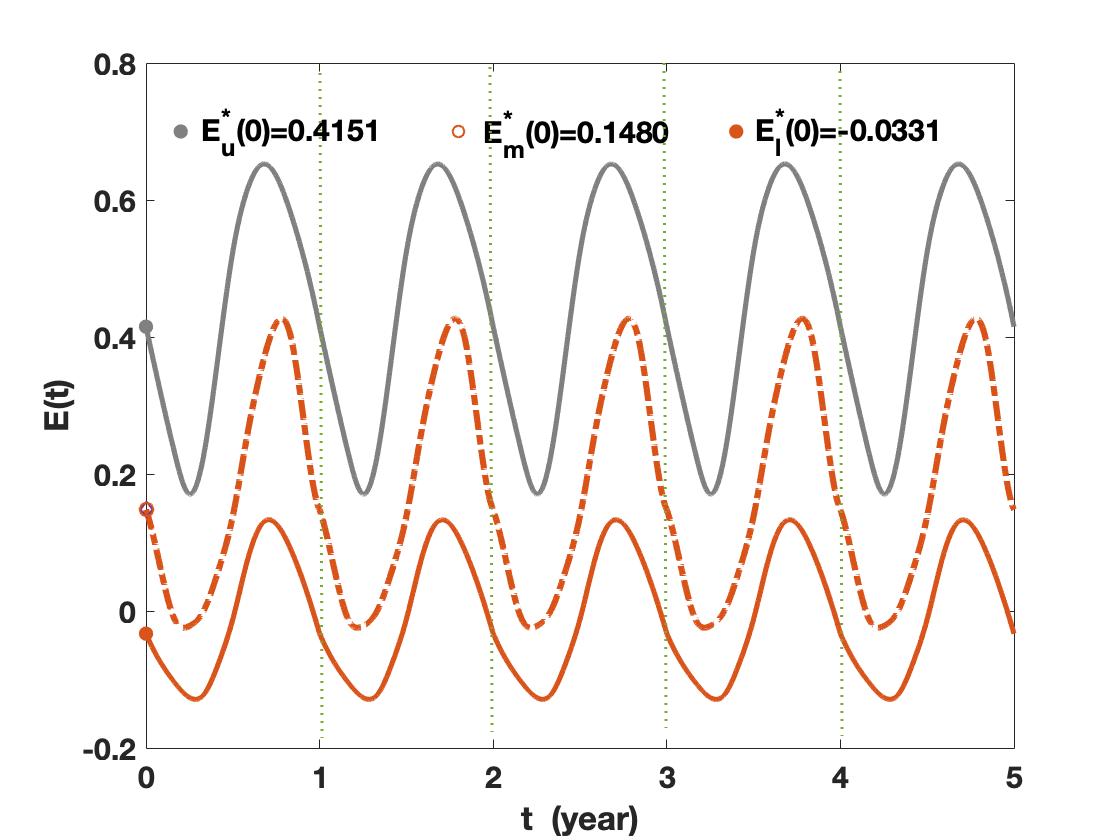}
	}
	\caption{The periodic solutions of the deterministic system $(\ref{e4})$ with different greenhouse gas forcing (a) $\Delta F_0=19$ and (b) $\Delta F_0=21$. A blue curve indicates the state that is perennially ice-covered ($E<0$ throughout the seasonally cycle). A grey curve indicates the state  that is perennially ice-free ($E>0$ throughout the year). A red solid curve indicates the state that is seasonally ice-free ($E<0$ and $E>0$ at different phases in the seasonal cycle). A red dashed curve indicates the state that  is an unstable intermediate state in which the Arctic Ocean is partially covered by ice and absorbs just enough solar radiation such that it remains at the freezing temperature: adding a small amount of additional sea ice to this unstable state would lead to less solar absorption, cooling, and a further extended sea-ice cover. }
	\label{F1}
\end{figure}
For the deterministic Arctic sea ice model $(\ref{e4})$, Eisenman and Wettlaufer$\cite{Eisenman2009}$ have constructed a Poincar\'e map to obtain the periodic solutions, as shown in Fig. \ref{F1}. Throughout this paper, the  ``lower'' and ``upper'' stable periodic solutions of the deterministic system are denoted by $E^{*}_l, E^{*}_u$, respectively, and the ``middle'' metastable one is denoted by $E^{*}_m$. In order to facilitate the numerical calculation, we write the deterministic model using the scale transformation $E/100$.
\begin{figure}
	\centering
	\includegraphics[width=0.4\textwidth]{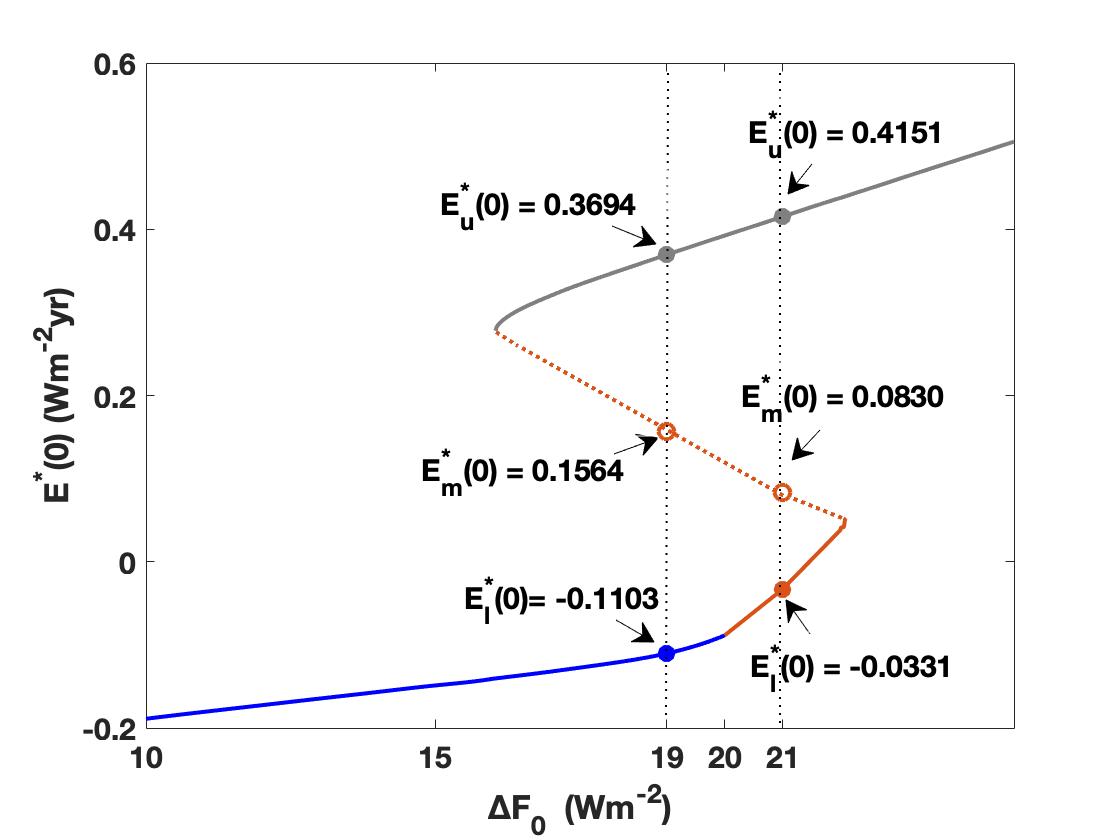}
	\caption{ A bifurcation diagram computed from a Poincar\'e map associated with equation (\ref{e4}). The bifurcation parameter is $\Delta F_0$ and the response of the system is represented by $E^{*}(0)$, the energy on 1 January.}
	\label{fig2}
	%\label{fig:subfig}%
\end{figure}

 In the case of Arctic sea ice,  a fixed point  corresponds to a steady state solution of the yearly cycle of ice growth and retreat \cite{Abbot2011}. In this paper we use the method used by Hill, Abbot, and Silber \cite{Hill2016} to construct Poincar\'e maps, which indicate the energy changes from 1 January in one year ($E_n(0)$) to 1 January in the next year ($E_{n+1}(0)=f(E_n(0))$) for a range of initial energies. Fixed points $E^{*}_{n}(0)$ of the Poincar\'e map occur when $E_{n+1}(0)=f(E^{*}_{n}(0))=E^{*}_{n}(0)$, and they correspond to the periodic solutions of the system with the same periodicity as the forcing (i.e., annual). For different greenhouse gas forcing, $\Delta F_0$, we use a Poincar\'e map to get the fixed points. As shown in Fig. $\ref{fig2}$, for $\Delta F_0 \in [10,16)$, there is only one stable fixed point,  corresponding to the perennially ice-covered state.
For $\Delta F_0 \in [16,20]$, there are two stable fixed points (the perennially ice-covered state and the perennially ice-free one) and one unstable intermediate fixed point. For $\Delta F_0 \in (20,23]$, there are two stable fixed points:  the seasonally ice-free state and the perennially ice-free state. For $\Delta F_0 \in (23,25]$, there is only one stable fixed point, which is the perennially ice-free state.

 \begin{figure*}[htbp]
 	\centering
 	
 	\subfigure[]
 	{
 		\label{fig3:subfig:a}%
 		\begin{minipage}[5]{.45\linewidth}
 			\centering
 			\includegraphics[width=1\textwidth]{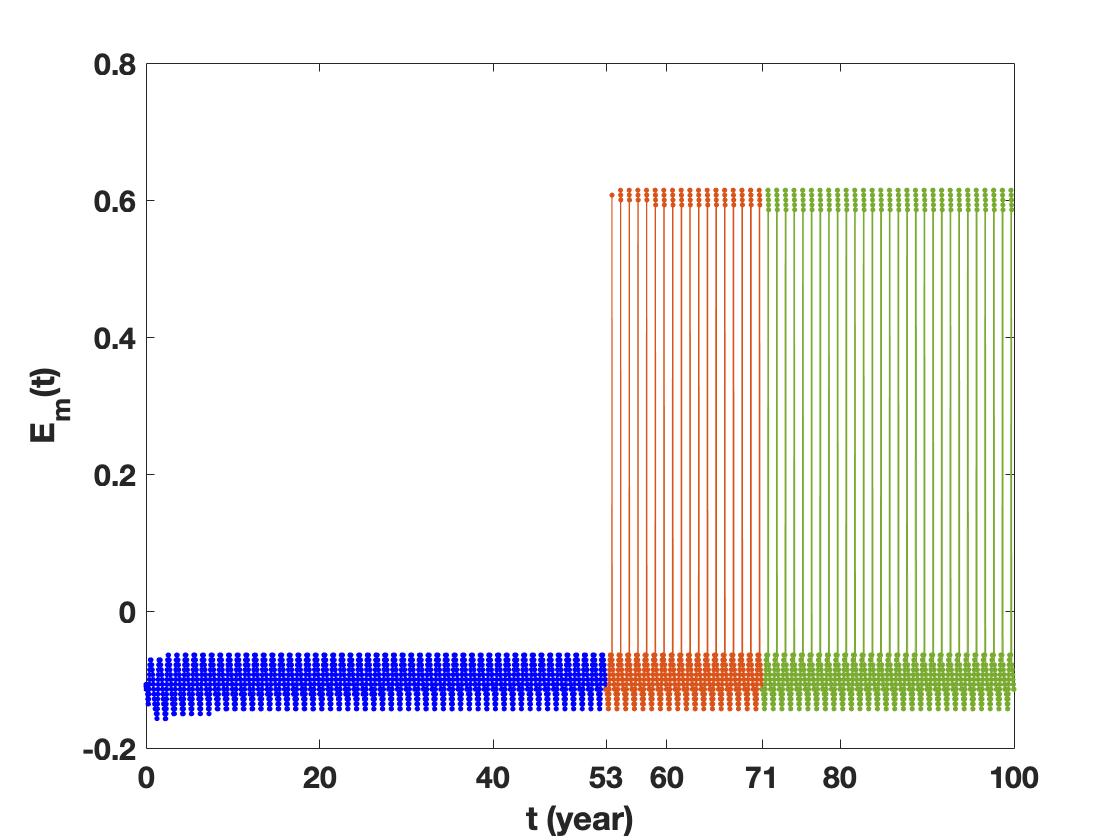}
 		\end{minipage}
 		
 	}	\subfigure[]
 	{
 		\label{fig3:subfig:b}%
 		\begin{minipage}[5]{.45\linewidth}
 			\centering
 			\includegraphics[width=1\textwidth]{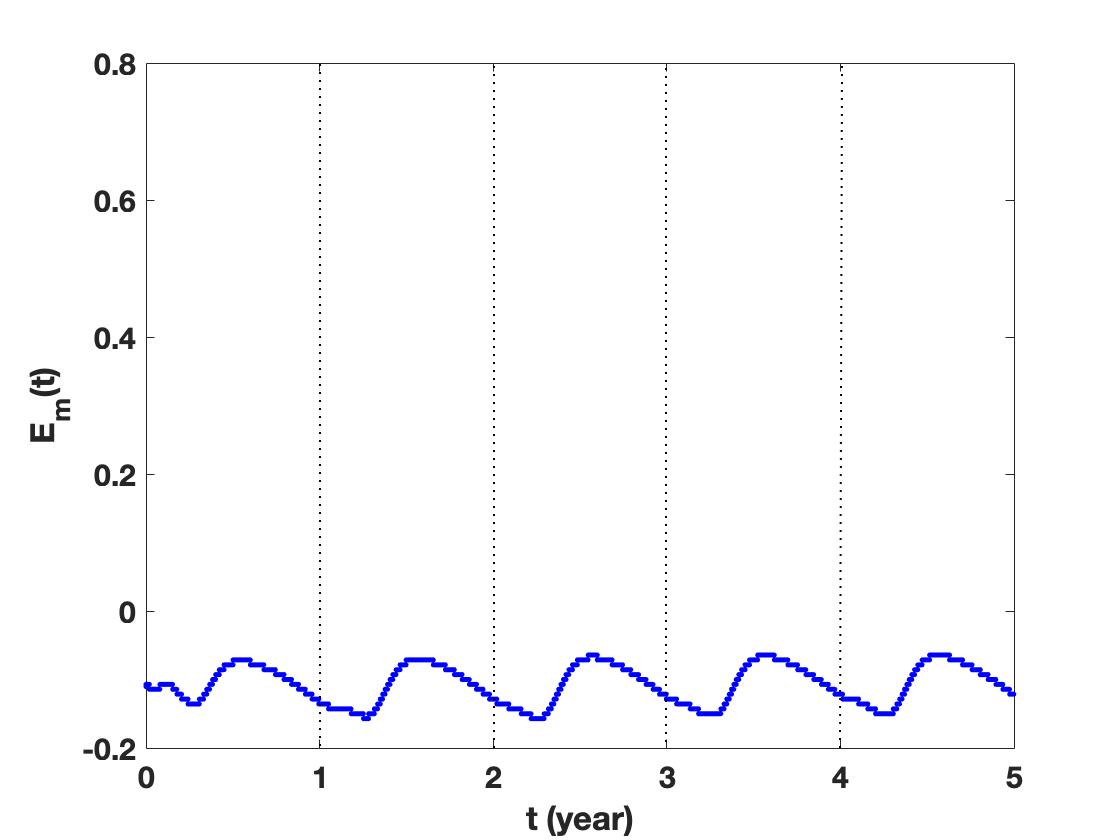}
 		\end{minipage}
 		
 	}	\subfigure[]
 	{
 		\label{fig3:subfig:c}%
 		\begin{minipage}[5]{.45\linewidth}
 			\centering
 			\includegraphics[width=1\textwidth]{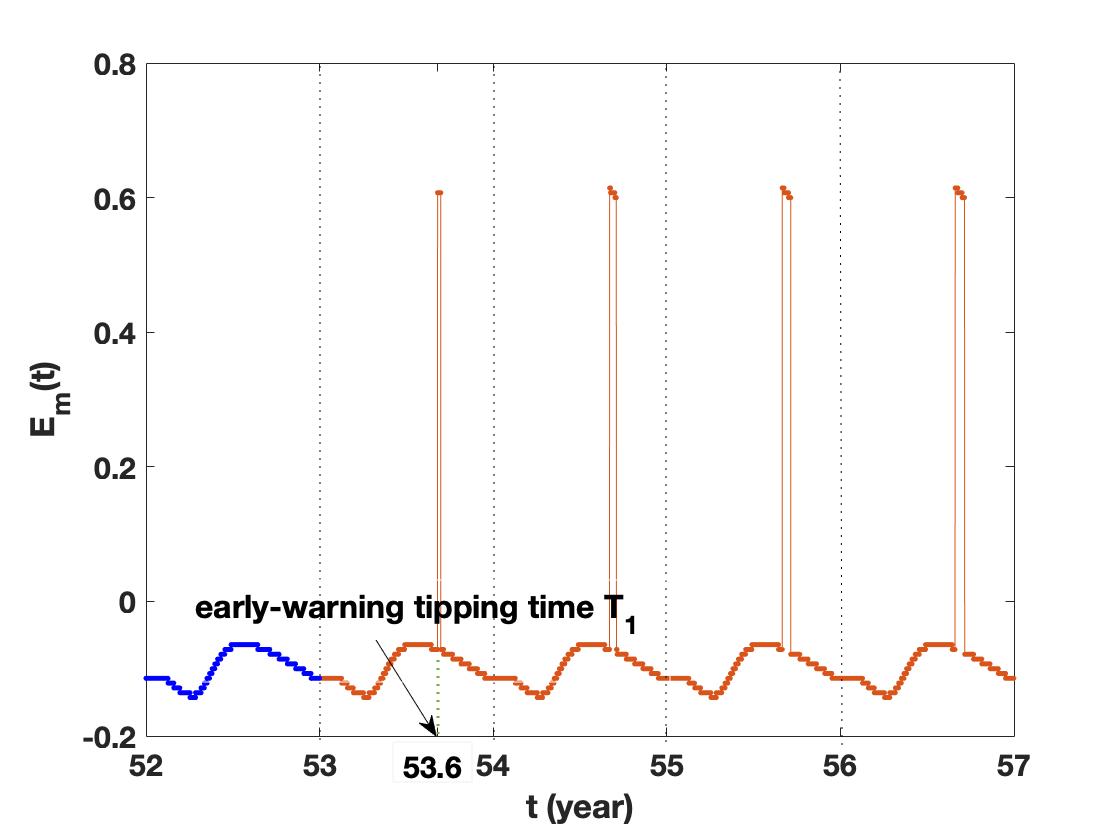}
 		\end{minipage}
 		
 	}	\subfigure[]
 	{
 		\label{fig3:subfig:d}%
 		\begin{minipage}[5]{.45\linewidth}
 			\centering
 			\includegraphics[width=1\textwidth]{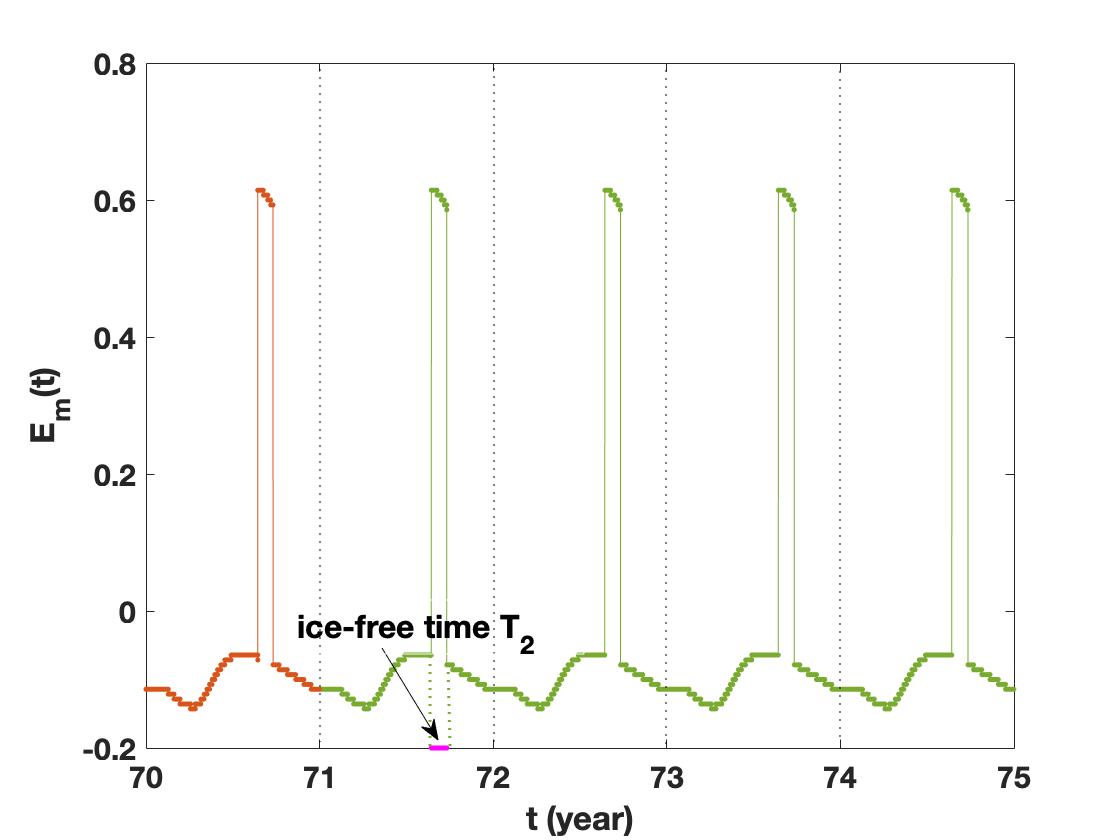}
 		\end{minipage}
 		
 	}
 	\caption{(a) The maximal likely trajectory of the nonautomous SDE $(\ref{e8})$ with the enhanced greenhouse gas level $\Delta F_0=19$, L\'evy index $\alpha =1.50$ and noise intensity $\epsilon =0.0450$. (b) The enlargement of the blue curve in (a), which shows a perennially ice-covered state ( since $E_m<0$ throughout the year).  (c) The enlargement of the connection between blue curve and red curve in (a), which displays the transition from the perennially ice-covered state to the seasonally ice-free state. (d) The enlargement of the connection between red curve and green curve in (a),  which shows that the system reaches the seasonally ice-free state.
 	}
 	
 \end{figure*}  	\label{fig3}
 %%%%%%%%%%%%%%%%%%%%%%%%%%%%%%%%%%%%%%%%%%%%%%
 \bigskip
 \textbf {D. Stochastic Arctic sea ice model}
 \bigskip

 The dynamical system $(\ref{e4})$ is a deterministic model. Even though it successfully captures the seasonal cycle of Arctic sea ice thickness and predicts the nature of the transitions as the greenhouse gas forcing $\Delta F_0$ increases. However, a realistic feature of the dynamic behavior of the ice cover is its variability due to internal fluctuations and external forcing \cite{Fetterer1998}.  Following Ditlevsen \cite{Ditlevsen1999}, we consider the following nonautonomous differential equation driven by a scalar symmetric $\alpha$-stable L\'evy process with  $1\leqslant \alpha \leqslant 2$:

 \begin{equation}
 dE=f(E,t)dt+\varepsilon dL^{\alpha}_t,
 \label{e8}
 \end{equation}
 where the term is $f(E,t)=\left(1-\alpha (E)\right)F_{S}(t)-F_{0}(t)-F_{T}(t)T(t,E)+\Delta F_{0}+F_{B}+\nu_{0} \mathit{R}(-E)$ corresponds to the right side of equation (5).  The positive quantity $\varepsilon $ denotes  the noise intensity. Thus, equation (\ref{e8}) represents the Arctic sea ice model subject to extreme events as modeled by a non-Gaussian $\alpha$-stable L\'evy process. This is a pure jump process when the L\'evy index $\alpha$ satisfies $1\leq \alpha <2$. It is known that a pure $\alpha$-stable L\'evy process has smaller jumps with higher jump probabilities as $\alpha$ approaches $2$.  In this stochastic system, the noise intensity and L\'evy index can be regarded as the intensity and the frequency of the extreme events.
The special case $\alpha=2$ corresponds to the usual  Gaussian process,  which is used to model ``normal" atmospheric fluctuations. 

 %%%%%%%%%%%%%%%%%%%%%%%%%%%%%%%%%%%%%%%%%%%%%

\section{Results\label{sec:level3}}

In this section, we analyse how the  noise intensity $\epsilon$, L\'evy index $\alpha$ and greenhouse gas forcing $\Delta F_0$ affect the maximal likely trajectory of the nonautonomous SDE $(\ref{e8})$. We determine the tipping times for transitions from the perennially ice-covered state to the seasonally ice-free one, and from the seasonally ice-free state to the perennially ice-free one. In the following, we choose one century (i.e. $T=100$) as the computational terminal time. Computations are carried out over a time interval of one century, that is, $T=100$ years. Following Eisenman and Wetlafer \cite{Eisenman2009} we use the terminology ``summer'' sea ice to refer to the annual minimum of Arctic sea-ice area and thickness and ``winter'' sea ice to refer to the annual maximum.

\bigskip

\textbf{A. Effect of $\alpha$-stable L\'evy process for the weakened greenhouse effect level $\Delta F_0=19$}

\medskip

For the relatively weak greenhouse gas level $\Delta F_0=19$, Fig. $\ref{fig3:subfig:a}$ illustrates the maximal likely trajectory  from the perennially ice-covered state to the seasonally ice-free state for L\'evy index $\alpha =1.50$ and noise intensity $\epsilon =0.0450$.
The blue curve indicates that the stochastic system maintains a perennially ice-covered state, which is enlarged in Fig. $\ref{fig3:subfig:b}$. The red curve indicates that the system has entered an unstable regime of operation. During the computational time interval of one century, the system begins to experience an ice-free state for a small fraction of time each year. Each year thereafter, the length of time in the fractionally ice-free state increases (see Fig. $\ref{fig3:subfig:c}$) until the system enters a stable regime of operation in which the length of time in the fractionally ice-free state remains constant, as shown in Fig. $\ref{fig3:subfig:d}$.

\medskip
 Based on the behavior discussed in the previous paragraph, we refer to the time when the system makes a first transition from a perennially ice-covered state to a state in which a fraction of each year is spent in an ice-free state as the \textit{early-warning tipping time}, and denote it by $T_1$. After $T_1$, the ice-covered period is decreasing until the Arctic appears in ice-free state in summer. $T_1$ could be regarded as an  early-warning signal of anomalous Arctic sea ice, and it  helps us to predict the approximate time when Arctic sea ice will disappear in summer.  If $T_1 = 100$, the system never appears ice-free and it remains in a perennially ice-covered state during our simulation cycle (one century).

% the stochastic Arctic sea ice system transfers from the stable perennial ice-covered state (seasonal ice-free state) to unstable seasonal ice-free state as shown in Fig. $\ref{fig3:subfig:c} $. \textcolor{red}{It means that the Arctic region most probable becomes ice-free during summertime at $T_1$.}

\begin{figure}[htbp]
	\centering
	\subfigure[]
	{
		\label{fig4:subfig:a}%
		\begin{minipage}[8]{.8\linewidth}
			\centering
			\includegraphics[width=1.1\textwidth]{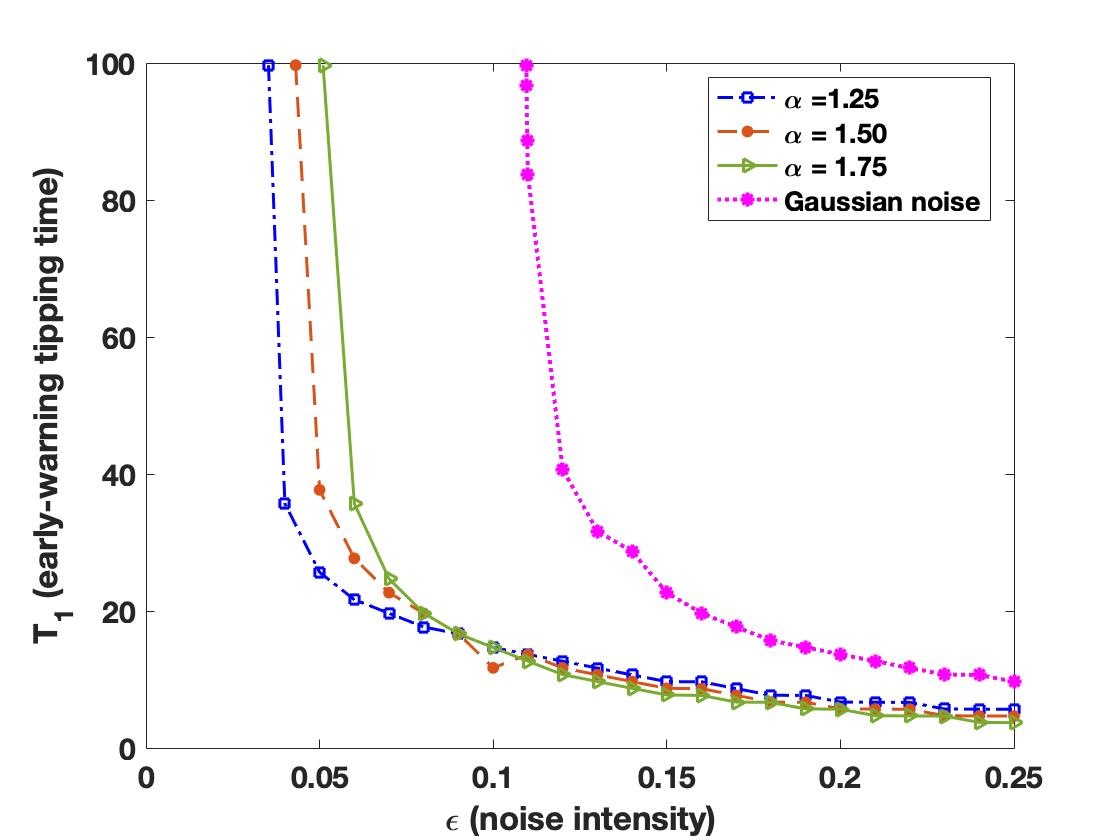}
		\end{minipage}
		
	}
	\subfigure[]
	{
		\label{fig4:subfig:b}%
		\begin{minipage}[8]{.8\linewidth}
			\centering
			\includegraphics[width=1.1\textwidth]{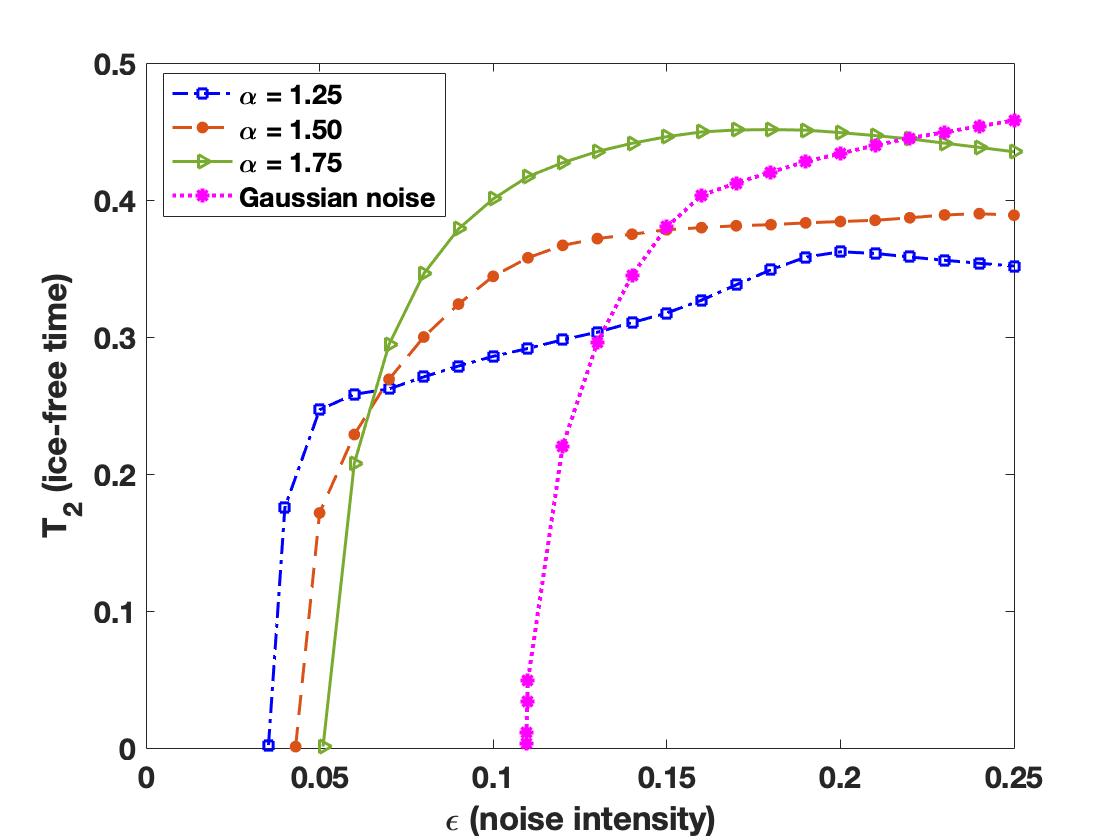}
		\end{minipage}
		
	}	
	\caption{Dependence of (a) the early-warning tipping time $T_1$ and (b) the ice-free time $T_2$ on the noise intensity $\epsilon$ for $\Delta F_0=19$ with non-Gaussian L\'evy noise  $\alpha=1.25, 1.5, 1.75$ and Gaussian noise.
}
	\label{F2}
\end{figure}

  We wish to understand the qualitative behavior of the dependence of the early-warning tipping time $T_1$ on the L\'evy index $\alpha \in [1,2]$ and the noise intensity $\epsilon$. This dependence is shown by the graphs in Fig. $\ref{fig4:subfig:a}$. We find that $T_1$ decreases as $\epsilon$ increases. This agrees with the corresponding result for Gaussian noise. For $\epsilon \in [0.051, 0.09]$, $T_1$ increases as $\alpha$ increases. However, when $\epsilon$ is greater than $0.09$, $T_1$ is insensitive to changes in $\alpha$. To determine the smallest value of $\epsilon$ that leads to a transition away from a perennially ice-covered regime within one century, draw a horizontal line in Fig. $\ref{fig4:subfig:a}$ at a level slightly below $T_1=100$. We see that the minimum noise intensity required for such a transition increases as $\alpha$ increases. For $\alpha=1.25, 1.5, 1.75$, the minimum values of $\epsilon$ are approximately $0.03528, 0.043096, 0.051$, respectively. For Gaussian noise, the minimum required noise intensity is clearly much larger.

We know that the $\alpha$-stable  L\'evy process has smaller jumps with higher jump probabilities for larger values of $\alpha$ ($1 \leqslant \alpha < 2$). This means that the frequency of extreme events increases when  L\'evy index $\alpha$ approaches $2$, and the intensity of extreme weather events increases as L\'evy noise intensity increases. These results on early-warning tipping times show that increased extreme events intensity lead to the earlier melting of the  sea ice.  For the small intensity of extreme events, the increased frequency of extreme events  has a positive effect on predicting the time for sea ice to melt in summer.

  \medskip

Once the sea ice begin to melt in summer, the system becomes unstable, and with evolution of a fraction of each year, the system will reach the seasonally ice-free state. In order to study the effect of non-Gaussian $\alpha$-stable noise on the time when ice-free state appeared in summer, we will introduce the \emph{ice-free time $T_2$} to represent the period of the year when the Arctic Ocean is ice-free as shown in Fig. $\ref{fig3:subfig:d}$.  It means that the larger $T_2$ is, the longer the a period of time for ice-free state in one year. If the climate becomes warmer further, it will be easier to appear a perennially ice-free state for the Arctic Ocean. It is worth pointing out that  the Arctic sea ice system remains in the ice-covered state all year round if the ice-free time $T_2=0$.

 Fig. $\ref{fig4:subfig:b}$ demonstrates the dependence of the ice-free time $T_2$ on the L\'evy index $\alpha \in [1, 2)$ and  the noise intensity $\epsilon$. We find that $T_2$ becomes longer as $\epsilon$ increases under influence of both Gaussian and non-Gaussian L\'evy noise. Another observation is that there is an intersection point near $\epsilon = 0.065$. When $\epsilon$ is larger thane the intersection point,  the ice-free time is longer as the value of $\alpha$ increases. However, the  ice-free time $T_2$ has opposite  behavior  when $\epsilon$ is smaller than $\epsilon = 0.065$.

 These results imply that one has to consider both the value of $\alpha$ and $\epsilon$ when we examine a fraction time of each year in an ice-free state. The ice-free period increases as the intensity of extreme events increases.  Meanwhile, we see that  the decreasing intensity and frequency of extreme events will be effective in reducing ice free times in summer. These results show the importance for studying extreme  weather events.

\begin{figure*}[htbp]
	\centering
	
	\subfigure[]
	{
		\label{fig6:subfig:a}%
		\begin{minipage}[8]{.45\linewidth}
			\centering
			\includegraphics[width=1.1\textwidth]{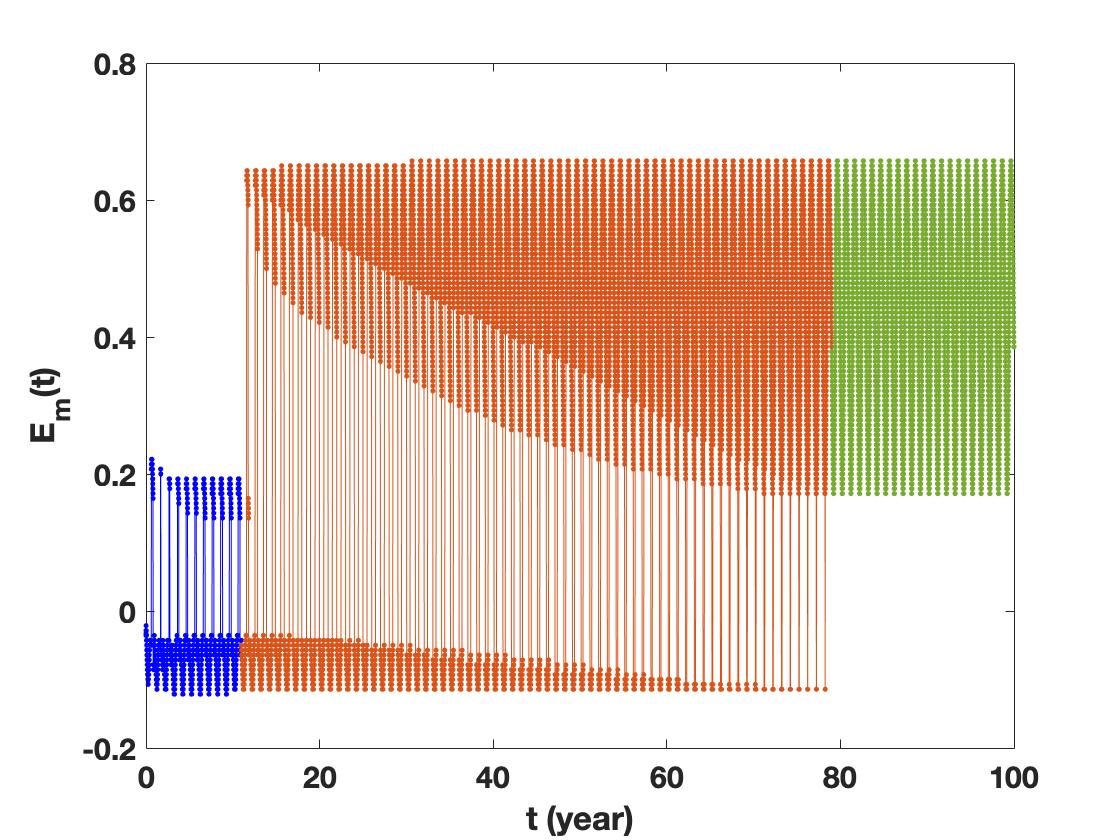}
		\end{minipage}
		
	}	\subfigure[]
	{
		\label{fig6:subfig:b}%
		\begin{minipage}[8]{.45\linewidth}
			\centering
			\includegraphics[width=1.1\textwidth]{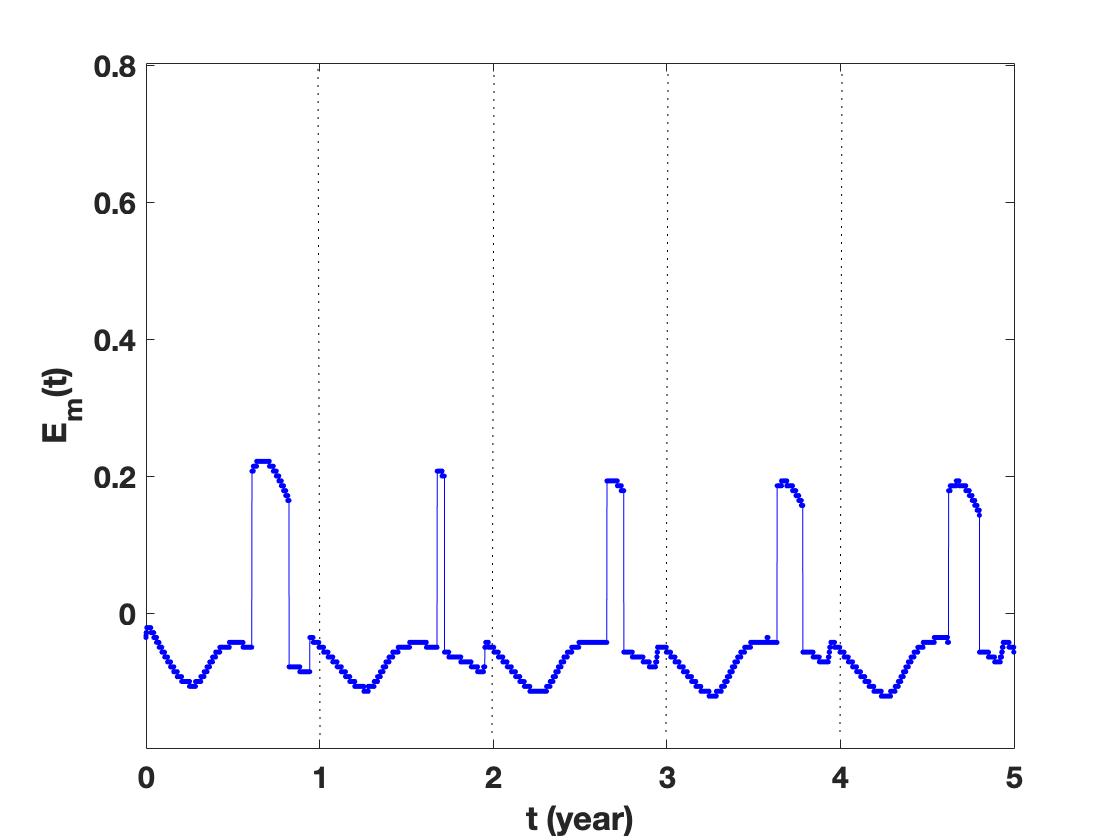}
		\end{minipage}
		
	}	\subfigure[]
	{
		\label{fig6:subfig:c}%
		\begin{minipage}[5]{.45\linewidth}
			\centering
			\includegraphics[width=1.1\textwidth]{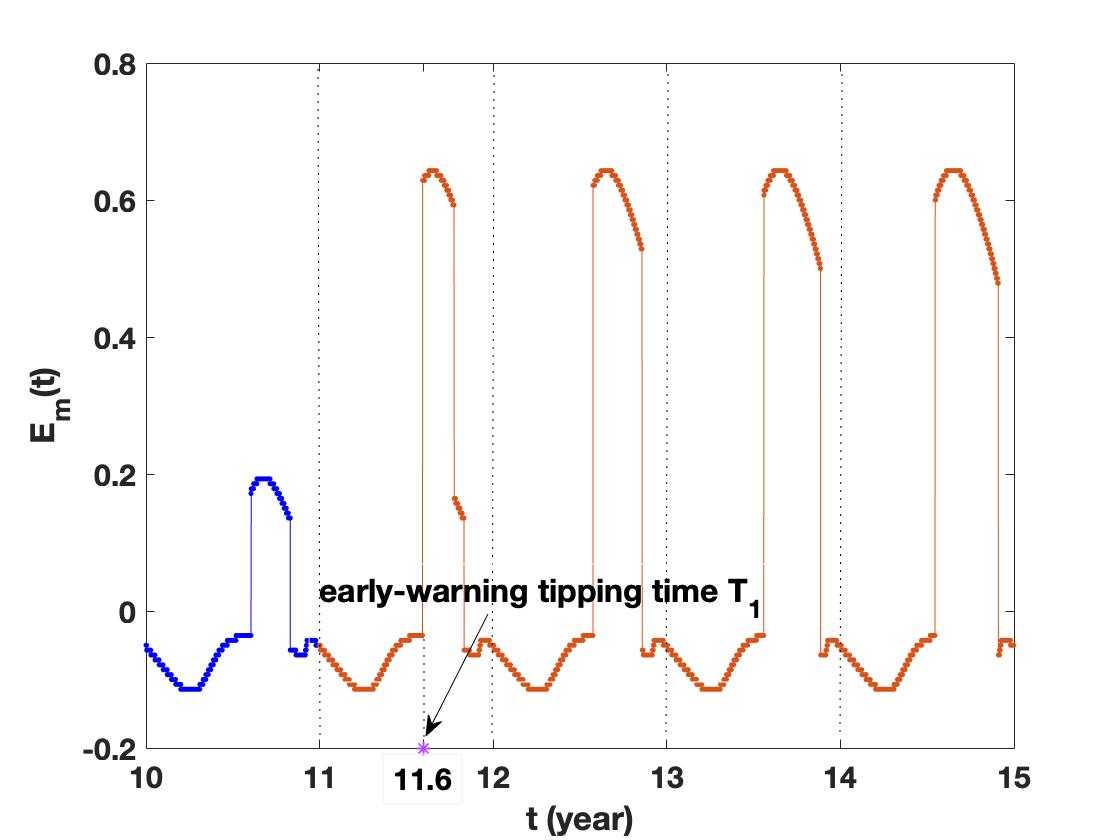}
		\end{minipage}
		
	}	\subfigure[]
	{
		\label{fig6:subfig:d}%
		\begin{minipage}[5]{.45\linewidth}
			\centering
			\includegraphics[width=1.1\textwidth]{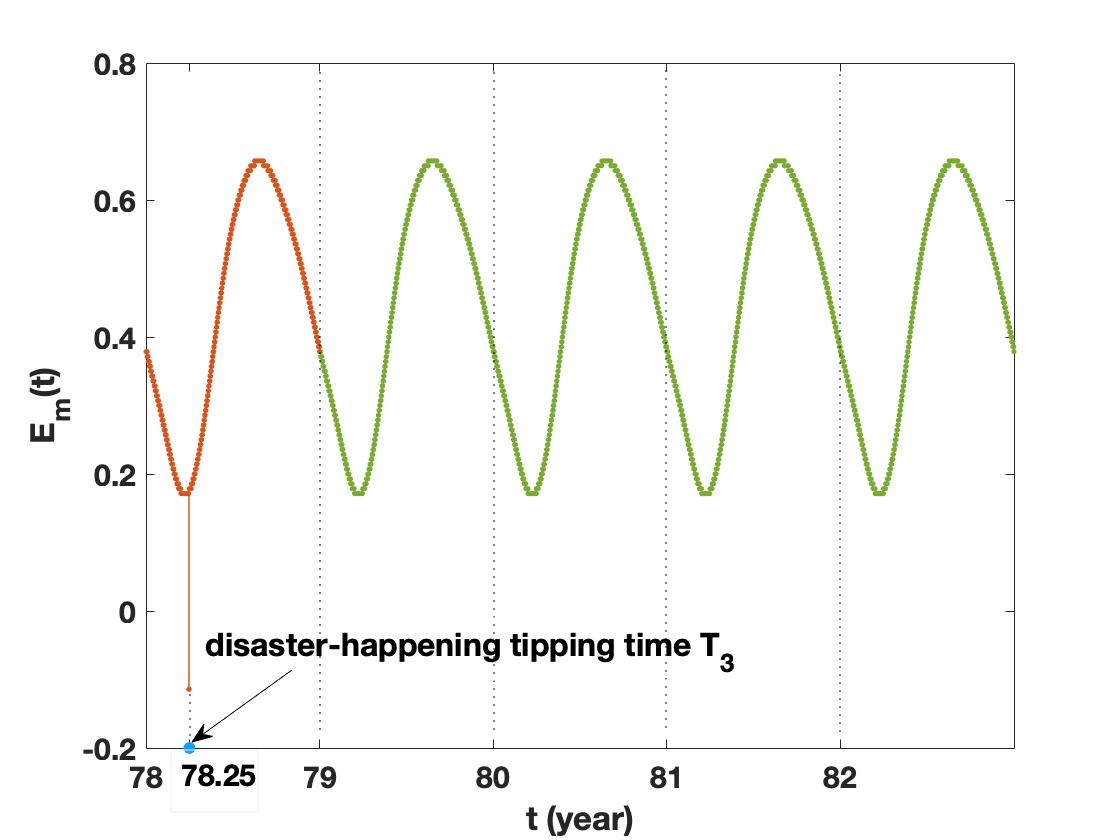}
		\end{minipage}
		
	}
	\caption{(a) The maximal likely trajectory of the SDE $(\ref{e8})$ with $\Delta F_0=21$, L\'evy index $\alpha =1.75$ and noise intensity $\epsilon =0.0250$. (b) The enlargement of blue curve in (a), which shows the seasonally ice-free state. (c) The enlargement of the connection between blue curve and red curve in (a), which displays the transition from the seasonally ice-covered state
to the  perennially ice-free state. (d) The enlargement of the connection between the red curve and the green curve in (a), which shows that the system reaches the perennially ice free state.
	}
	\label{F2}
\end{figure*}
\bigskip

\begin{figure}[htbp]
	\centering
	\subfigure[]
	{
		\label{fig7:subfig:a}%
		\begin{minipage}[5]{.8\linewidth}
			\centering
			\includegraphics[width=1.1\textwidth]{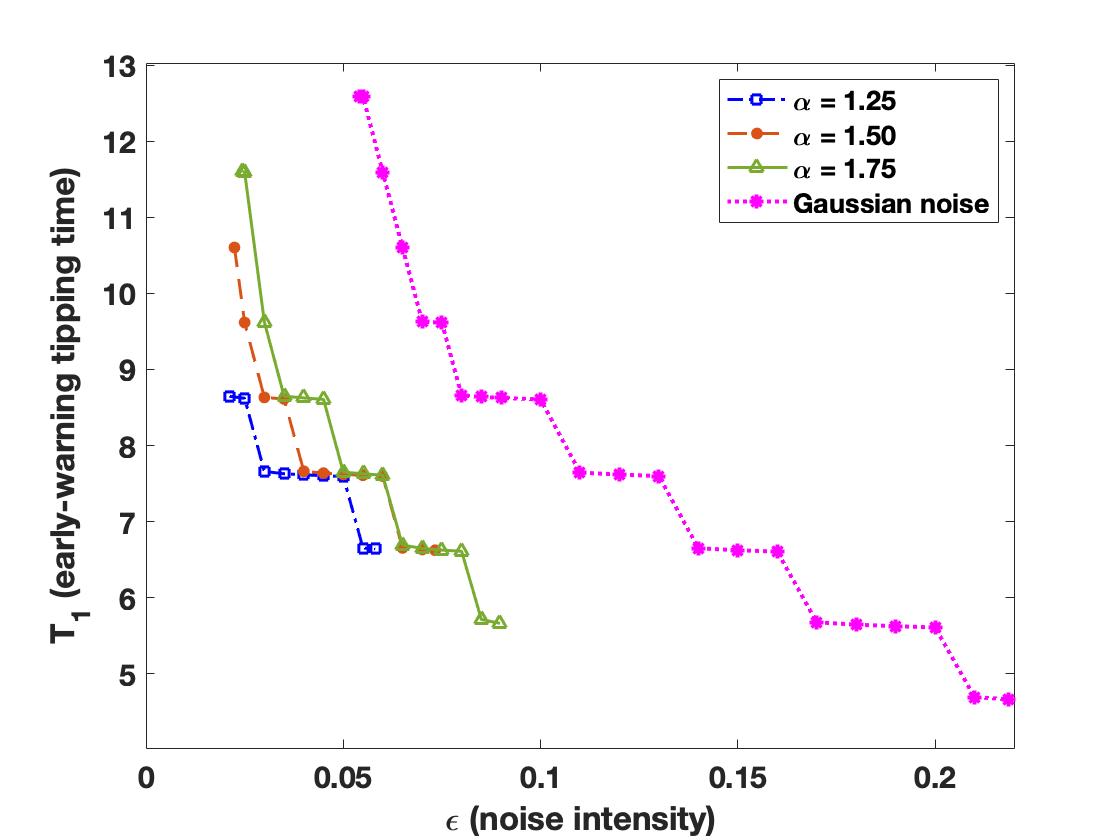}
		\end{minipage}
		
	}

	\subfigure[]
{
	\label{fig7:subfig:b}%
	\begin{minipage}[5]{.8\linewidth}
		\centering
		\includegraphics[width=1.1\textwidth]{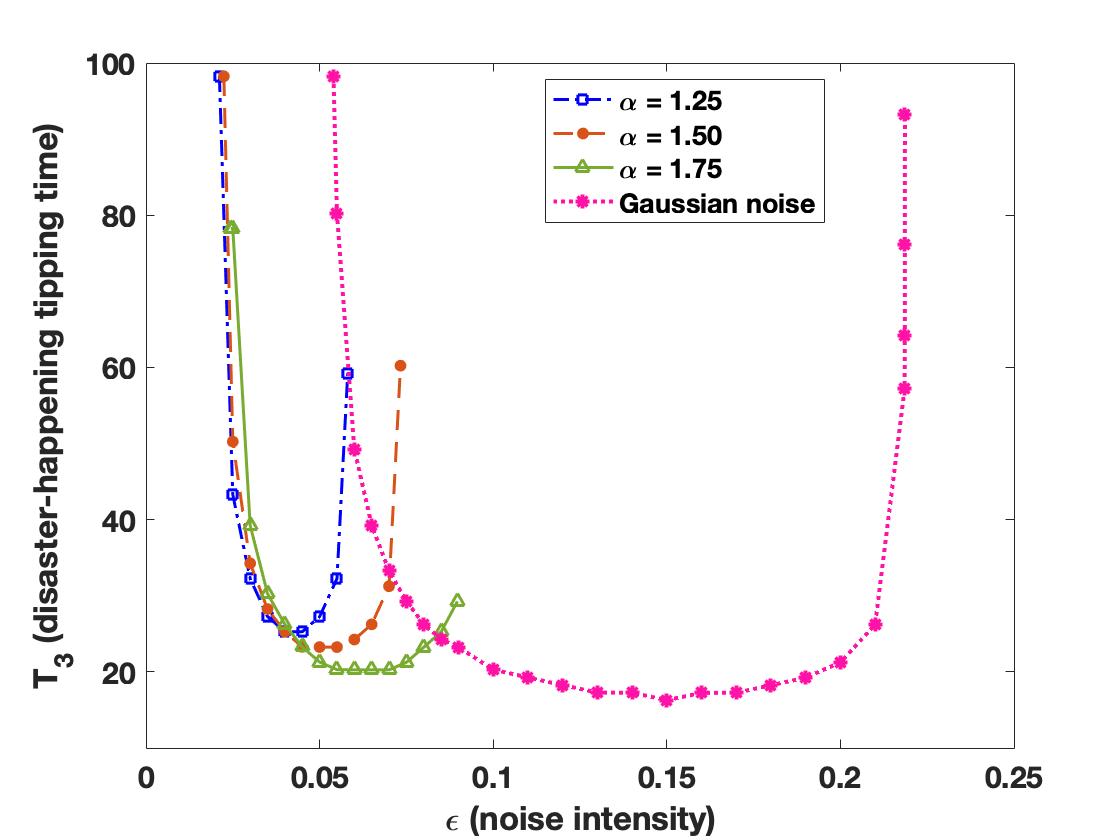}
	\end{minipage}
	
}
	
	\caption{ Dependence of (a) the early-warning tipping time $T_1$ and (b) the disaster-happening time $T_3$ on the noise intensity $\epsilon$ for $\Delta F_0=21$ with non-Gaussian L\'evy noise $\alpha=1.25, 1.5, 1.75$ and Gaussian noise.}
	\label{F2}
\end{figure}

\bigskip
\textbf{B. Effect of $\alpha$-stable L\'evy process for the enhanced greenhouse effect level $\Delta F_0=21$}

\begin{figure}[htbp]
	\centering
	\subfigure[]
	{
		\label{fig8:subfig:a}%
		\begin{minipage}[8]{.8\linewidth}
			\centering
			\includegraphics[width=1.1\textwidth]{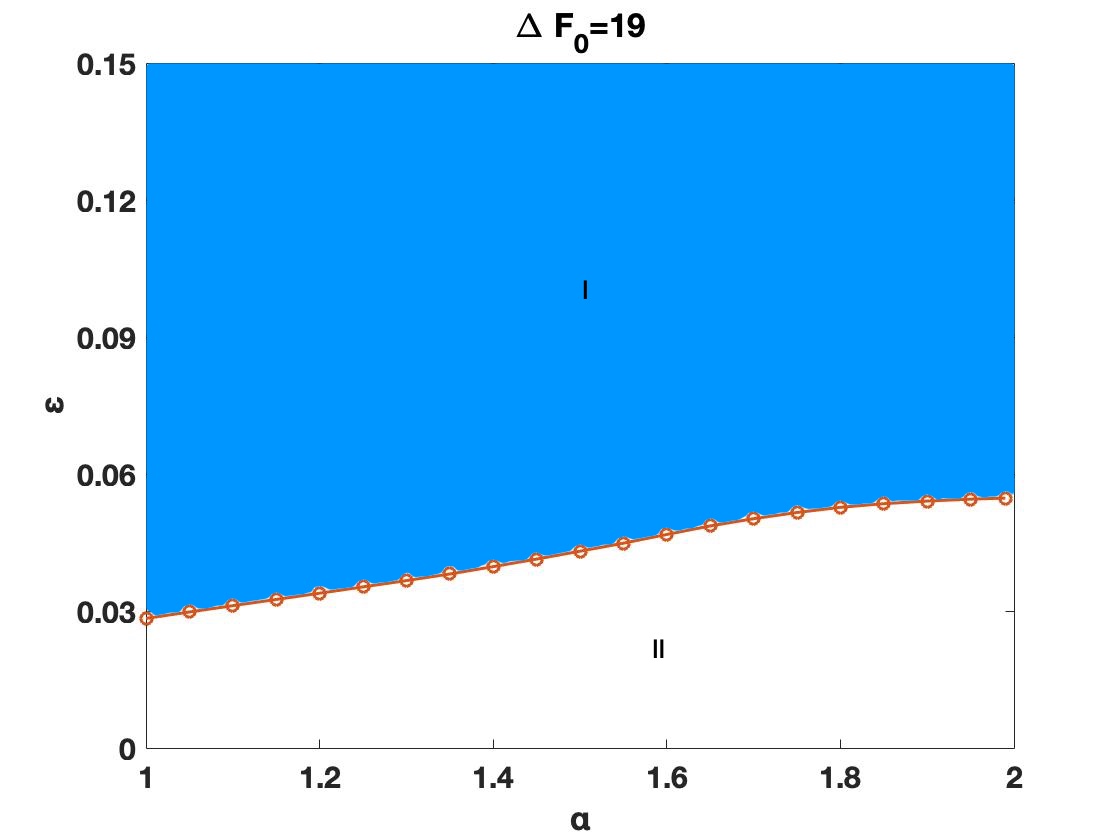}
		\end{minipage}
		
	}
	\subfigure[]
	{
		\label{fig8:subfig:b}%
		\begin{minipage}[8]{.8\linewidth}
			\centering
			\includegraphics[width=1.1\textwidth]{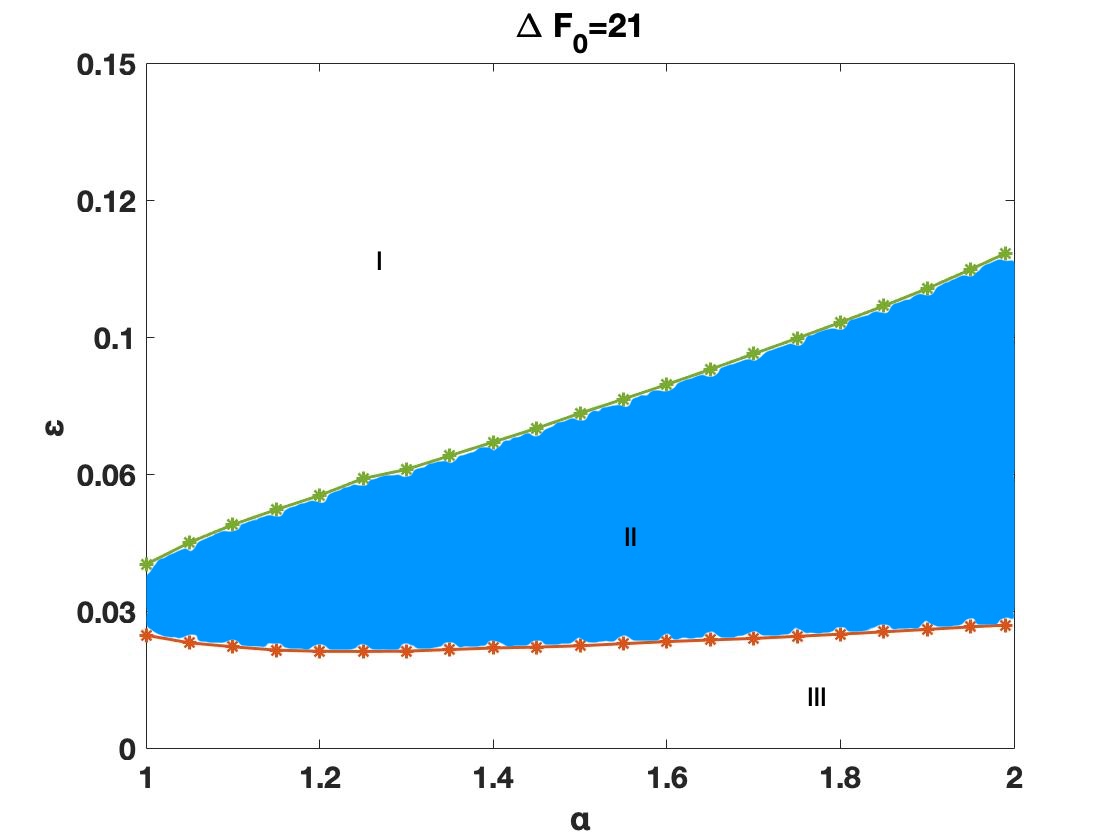}
		\end{minipage}
		
	}	
	\caption{The combination of $\epsilon$ and $\alpha$ which can trigger the transition between metastable states in the Arctic sea ice system. (a) For $\Delta F_0=19$, region I: transition from the perennially ice-covered state to the seasonally ice-free one (b) For $\Delta F_0=21$, region II: the transition of the Arctic sea ice evolve from the seasonally ice-free state to the perennially ice-free state.}
	\label{F2}
\end{figure}
\medskip

For this larger value of $\Delta F_0 = 21 $, the maximal likely trajectory of the nonautonomous SDE $(\ref{e8})$ from the seasonally ice-free state to the perennially ice-free state for L\'evy index $\alpha =1.75$ and noise intensity $\epsilon =0.0250$ is shown in Fig. $\ref{fig6:subfig:a}$. The blue curve represents the stochastic system concentrated on the seasonally ice-free state, as enlarged in Fig. $\ref{fig6:subfig:b}$. The red curve  means that the system has experienced an unstable ice-free state, the length of time in the fractionally ice-free state increases (Fig. $\ref{fig6:subfig:c}$) until the system remains in a perennially ice-free metastable states (as $E_m > 0$ throughout the year), as shown in  Fig. $\ref{fig6:subfig:d}$.

\medskip

To analyze how the system of Arctic sea ice shifts from the seasonally ice-free state to the perennially ice-free state under noise, we will continue to use the early-warning tipping time $T_1$ to denote the first time when the seasonally ice-free state changes, as shown in Fig. \ref{fig6:subfig:c}. After $T_1$, a fraction of ice-free time increases until the winter sea ice vanishes to the perennially ice-free state. Once this time $T_1$ is reached, we should develop and deploy adaptive strategies, and take a more pre-emptive, precautionary policy approach to prevent the situation from getting worse.

Fig. \ref{fig7:subfig:a} shows that the early-warning tipping time $T_1$ presents a ladder descending trend with the increased  noise intensity $\epsilon$ for non-Gaussian L\'evy noise with  $\alpha=1.25,~1.50,~1.75$.  Additionally, for different $\alpha$, we find that the range of L\'evy noise intensity that enables the system to shift to the perennially ice-free state is different.  For example, for $\alpha=1.25, 1.75$, the corresponding transition range of $\epsilon$ is $[0.0211, 0.058]$ and $[0.0245, 0.0897]$, respectively. On the other hand, for the Gaussian noise, the behaviour of $T_1$ agrees in general with the results in previous pure jump case. However, compared with non-Gaussian L\'evy noise,  Gaussian noise requires the stronger noise intensity and have the larger range of noise intensity for transition to the seasonally ice-free state.

 Furthermore, we find that $T_1$ obviously appears earlier for the enhanced greenhouse level $\Delta F_0=21$ by comparing with the weakened greenhouse level $\Delta F_0=19$, as shown in Fig. \ref{fig7:subfig:a} and \ref{fig4:subfig:a}. For example, keeping $\alpha=1.25$ and $\epsilon=0.04$, the early-warning tipping time close to $35$ and $7$ for $\Delta F_0=19$ and $21$, respectively. This means that the Arctic sea ice will melt much earlier under influence of the enhanced greenhouse effect.

For $\Delta F_0=21$, the ice-free time  $T_2 \equiv 1$, because the system keep in the perennially ice-free state. In this case, the Arctic Ocean is ice-free all over the year. Therefore, the ice-free time $T_2$ can not capture the dynamical behaviour in this case.  Next, we will introduce the other quantity, which can help us to predict the approximate time when the thickest ice sheet of the Arctic region in a year will disappear. The \emph{disaster-happening tipping time T3} will be introduced to express the last time when the Arctic Ocean has the winter sea ice, and after that time, the Arctic Ocean remains ice-free throughout the year, as shown in Fig. \ref{fig6:subfig:d}. The time $T_3$ is a signal of disaster that may happen in Arctic sea ice with unimaginable consequences--from the loss of polar bear habitat to possible increases of weather extremes at mid-latitudes \cite{ Kerr2012}. We hope that the $T_3$ never shows up.

 Fig. \ref{fig7:subfig:b} shows the numerical results of the disaster-happening tipping time $T_3$ effect of the noise intensity $\epsilon$  and L\'evy index $\alpha=1.25,~1.50,~1.75$. We find that $T_3$ shows a U-shaped changing trend with the increasing of value of $\epsilon$ for the case of non-Gaussian L\'evy noise. In the beginning, $T_3$ decreases rapidly as $\epsilon$ increases. Then, the effect of the increasing $\epsilon$ on $T_3$ is relatively weak. Finally, $T_3$ will turn a corner when the curve reaches a nadir, after this inflection point, $T_3$ rapidly grows with increasing of $\epsilon$. Furthermore, the value of $\alpha$ is larger, the growth rate of $T_3$ is  slower.  This means that it will delay the time for the appearance of the perennially ice-free state when the  intensity of extreme events increases to a certain degree. Meanwhile, the stronger the frequency of extreme events is, the slower the delay time will be. A possible explanation for these inflection points could be the interaction between the ice albedo feedback and the greenhouse effect. We note that for the  case of Gaussian noise, relations between the disaster-happening tipping time $T_3$ and the value of noise intensity are similar to the non-Gaussian L\'evy noise, while the appearance of inflection point requires the stronger noise intensity.

% The early-warning time $T_1$ can be simply interpreted as the time at which the thinnest ice sheet of the Arctic in summer during one whole year most probably disappear. The disaster-happening time $T_3$ can be simply interpreted as the time when the thickest ice sheet of the Arctic in a whole year most probably disappear. For

%%%%%%%%%%%%%%%%%%%%%%%%%%%%%%%%%%%%%%%%%%%%%%
\bigskip
\textbf{C. Combination of L\'evy noise intensity $\epsilon$ and L\'evy index $\alpha$ trigger transition}
\medskip

For $\Delta F_0=19$ and a fixed value of  $\alpha \in [1,2)$,  Fig. \ref{fig8:subfig:a} shows the minimum  $\epsilon$ that can trigger transition  from the perennially ice-covered state to the seasonally ice-free state, which is marked with red solid points. For example, for $\alpha = 1.0,1.75$, the minimum noise intensity that can trigger the transition is $\epsilon \approx 0.03, 0.051$, respectively. The larger $\alpha$ is, the larger noise intensity is needed. We collect all the combinations of $\alpha$ and $\epsilon$  that can induce the transitions for Arctic sea ice system in  the region I .

Similarly, the combinations of $\alpha$ and $\epsilon$ leading to the transition from the seasonally ice-free state to the perennially ice-free state can be obtained for $\Delta F_0=21$, as shown in Fig. \ref{fig8:subfig:b}. For example, for $\alpha =1.5$, the minimum value of noise intensity is $\epsilon \approx 0.02114 $ and the maximum value of noise intensity is  $\epsilon \approx 0.055$. As shown in Fig. $\ref{fig8:subfig:b}$, the  blue curve and red curve denote an upper bound and a lower bound of $\epsilon$, which correspond to the maximum and minimum noise intensities inducing the transition, respectively. These two curves constitute the region II  where the system can shift to a perennially ice-free state from a seasonally ice-free state .

%%%%%%%%%%%%%%%%%%%%%%%%%%%%%%%%%%%%%%%%%%%%%%
\section{Conclusions \label{sec:level4}}

    Arctic sea ice in recent summers shows the record lows in ice extent \cite{Stroeve2005, Walker2016}.    It  has also been  thinning at a remarkable rate over the past few decades \cite{Bitz2004,Mueller2011}. To insight into the Arctic sea ice variations under influence of extreme weather events, we propose a nonautonomous  Arctic sea ice model under the non-Gaussian $\alpha$-stable L\'evy noise. We use the maximal likely trajectory,  based on the numerical solution of the nonlocal Fokker-Planck equation (\ref{e2}), to obtain the tipping times for the stochastic Arctic sea ice model (\ref{e8}). The early-warning tipping time $T_1$ and the disaster-happening tipping time $T_3$ are used to predict the time when the Arctic Ocean may appear in ice-free state in summer and  in winter, respectively.

By  numerical experiments, we find that the tipping times $T_1$ and $T_3$ depend strongly on the L\'evy index $\alpha$, noise intensity $\epsilon$ and greenhouse effect $\Delta F_0$. For example, for the enhanced greenhouse level $\Delta F_0=21$,  the  early-warning tipping time $T_1$ decreases with the increased intensity of $\alpha$-stable L\'evy noise, in agreement with the corresponding results for the weakened greenhouse level $\Delta F_0=19$, but $T_1$ comes earlier for $\delta F_0 = 21$. This implies that Arctic sea ice will melt much earlier in summer under influence of the enhanced greenhouse effect and increased noise intensity. On the other hand, the disaster-happening tipping time $T_3$ shows a U-shaped changing trend under non-Gaussian  L\'evy noise with the increase of value of $\epsilon$. Another observation from the results is that for the ice-free time $T_2$, we see that  the decreased intensity and frequency of extreme events will be effective in reducing the length of time in the fractionally ice-free state. 

Comparing with the non-Gaussian L\'evy noise, we discover that a larger noise intensity  is needed  to induce a transition between metastable states in Gaussian  case (see Fig.4 and Fig.6). Finally, we identify   the combinations of $\alpha $ and $\epsilon$ that trigger the transitions from one state to the other one in the Arctic sea ice system under non-Gaussian  L\'evy noise for both the weakened and the enhanced greenhouse gas levels. 

Our work  provides a theoretical framework for studying and predicting the Arctic sea ice variations  under the influence of extreme events.

\section*{Appendix A: symmetric $\alpha$-stable L\'evy process}

It is well known that Brownian motion has the stationary, the independent increments and almost surely continuous sample paths. A L\'evy process is a non-Gaussian process with heavy-tailed statistical distribution and intermittent bursts.  The stable distribution $S_{\alpha}(\delta, \beta, \lambda)$ is determined by the following four indexes: L\'evy index $\alpha \in (0,2)$, scale parameter $\delta \in [0, \infty)$,  skewness parameter $\beta \in [-1,1]$ and shift parameter $\lambda \in(-\infty, \infty)$  \cite{Duan2015}.

A scalar symmetric $\alpha$-stable L\'evy process $L^{\alpha}_{t}$ is defined  via the following properties:
\begin{itemize}
	\item[(i)] $L^{\alpha}_{0}=0$, almost surely (a.s.);
	\item[(ii)] $L^{\alpha}_{t}$ has independent increments: For every      natural number $n $ and positive  time instants with  $t_{0}<t_{1}< \cdots <t_{n}$, the  random variables   $L^{\alpha}_{t_{1}}-L^{\alpha}_{t_{0}}$, $L^{\alpha}_{t_{2}}-L^{\alpha}_{t_{1}}$, $\cdots$, $L^{\alpha}_{t_{n}}-L^{\alpha}_{t_{n-1}}$ are independent;
	\item[(iii)] $L^{\alpha}_{t}$ has stationary increments: $L^{\alpha}_{t}-L^{\alpha}_{s}$ and $L^{\alpha}_{t-s}$ have the same stable distribution $S_{\alpha}((t-s)^{\frac1{\alpha}}, 0, 0)$, for $t>s$;
	\item[(iv)] $L^{\alpha}_{t}$ has stochastically continuous sample paths:  $L^{\alpha}_{t} \rightarrow L^{\alpha}_{s}$ in probability, as $t \rightarrow s$.
\end{itemize}

An $\alpha$-stable L\'evy process $L^{\alpha}$ has the following ``heavy tail'' estimate \cite{Zheng2020}:

\begin{equation*}
\lim_{y \rightarrow \infty} y^{\alpha} \mathbb{P}(L^{\alpha}>y)=C_{\alpha}\frac{1+\beta}{2} \sigma^{\alpha},
\end{equation*}
i.e., the tail estimate decays in power law. The constant $C_{\alpha}$ is a gamma function depending on $\alpha$.

For a symmetric $\alpha$-stable L\'evy process $L^{\alpha}_{t}$, the skewness index $\beta$ is equal to $0$.  The paths of $L^{\alpha}_{t}$ have countable jumps, and the jump measure $\nu_{\alpha}(dy)$ depends on $\alpha$,
\begin{equation*}
\nu _{\alpha} (dy)=C_{\alpha} \mid y \mid ^{-(1+\alpha)} dy,
\end{equation*}
For $0<\alpha <1$, $L^{\alpha}_{t}$ has larger jumps with lower jump probabilities, while for $1<\alpha <2$, it has smaller jumps with higher jump frequencies. The special case for $\alpha =2$ corresponds to a Brownian motion, and it is a Cauchy process for $\alpha=1$. More information on $\alpha$-stable L\'evy process, see references $\cite{Saltzman2002, Duan2015}$.

\appendix
\section*{Appendix B: Descriptions and Default Values of Model Parameters}
\begin{table}[h]
	\tiny
	\caption{\label{tab:table1}Descriptions and Default Values of Model Parameters}
	\begin{ruledtabular}
		\begin{tabular}{lll}
			Symbol&Description& Value\\
			\hline
			$L_{i}$                & Latent heat of fusion of ice &9.5 $Wm^{-3}yr$\\
			$c_{ml}H_{ml}$ &Ocean mixed layer heat capacity times depth & $6.3Wm^{-2}yrK^{-1}$\\
			$k_{i}$                &Ice thermal conductivity& $2Wm^{-1}K^{-1}$\\
			$F_{B}$               &Heat flux into bottom of sea ice or ocean mixed layer & $2Wm^{-2}$\\
			$h_{\alpha}$       &Ice thickness range for smooth transition from $\alpha_{i}$ to $\alpha_{ml}$&$0.5m$\\
			$\nu_{0}$             &Dynamic export of ice from model domain& $0.1yr^{-1}$\\
			$\alpha_{i}$       &Albedo when surface is ice cover&$0.68$\\
			$\alpha_{ml}$    &Albedo when ocean mixed layer is exposed&$0.2$\\
			$F_{0}(t)$          &Temperature-independent surface flux (seasonally varying)&$85 Wm^{-2}$\\
			$F_{T}(t)$             &Temperature-dependent surface flux (seasonally varying)&$2.8 Wm^{-1}K^{-1}$\\
			$F_{S}(t)$  &Incident shortwave radiation flux (seasonally varying)&$100 Wm^{-2}$\\
			$\Delta F_{0}$ &Imposed surface heat flux& $0 Wm^{-2}$\\
		\end{tabular}
	\end{ruledtabular}
\end{table}
For the seasonally varying parameters $F_0(t)$, $ F_T(t)$ and $F_S(t)$, the  monthly values  starting with January are  $F_{0}(t)$=(120, 120, 130, 94, 64, 61, 57, 54, 56, 64, 82, 110) $Wm^{-2}$,
$F_{T}(t)$=(3.1, 3.2, 3.3, 2.9, 2.6, 2.6, 2.6, 2.5, 2.5, 2.6, 2.7, 3.1)$Wm^{-2}K^{-1}$, and $F_{S}(t)$=(0, 0, 30, 160, 280, 310, 220, 140, 59, 6.4, 0, 0)$Wm^{-2}$.

\bigskip
\setcounter{equation}{0}

\begin{acknowledgements}
The authors would like to thank Professor Xu Sun, Dr Xiaoli Chen, Dr Xiujun Cheng, and  Dr Yuanfei Huang for helpful discussions and comments. This work was partly supported by the   NSFC grants 11801192, 11531006 and 11771449.
\end{acknowledgements}

\section*{Data Availability Statement}
The data that support the findings of this study are openly available on GitHub \cite{Yang2020}.

\bigskip

%\nocite{*}
%\bibliographystyle{unsrt}
%\bibliography{maxlikehood}

\begin{thebibliography}{56}%
\makeatletter
\providecommand \@ifxundefined [1]{%
 \@ifx{#1\undefined}
}%
\providecommand \@ifnum [1]{%
 \ifnum #1\expandafter \@firstoftwo
 \else \expandafter \@secondoftwo
 \fi
}%
\providecommand \@ifx [1]{%
 \ifx #1\expandafter \@firstoftwo
 \else \expandafter \@secondoftwo
 \fi
}%
\providecommand \natexlab [1]{#1}%
\providecommand \enquote  [1]{``#1''}%
\providecommand \bibnamefont  [1]{#1}%
\providecommand \bibfnamefont [1]{#1}%
\providecommand \citenamefont [1]{#1}%
\providecommand \href@noop [0]{\@secondoftwo}%
\providecommand \href [0]{\begingroup \@sanitize@url \@href}%
\providecommand \@href[1]{\@@startlink{#1}\@@href}%
\providecommand \@@href[1]{\endgroup#1\@@endlink}%
\providecommand \@sanitize@url [0]{\catcode `\\12\catcode `\$12\catcode
  `\&12\catcode `\#12\catcode `\^12\catcode `\_12\catcode `\%12\relax}%
\providecommand \@@startlink[1]{}%
\providecommand \@@endlink[0]{}%
\providecommand \url  [0]{\begingroup\@sanitize@url \@url }%
\providecommand \@url [1]{\endgroup\@href {#1}{\urlprefix }}%
\providecommand \urlprefix  [0]{URL }%
\providecommand \Eprint [0]{\href }%
\providecommand \doibase [0]{http://dx.doi.org/}%
\providecommand \selectlanguage [0]{\@gobble}%
\providecommand \bibinfo  [0]{\@secondoftwo}%
\providecommand \bibfield  [0]{\@secondoftwo}%
\providecommand \translation [1]{[#1]}%
\providecommand \BibitemOpen [0]{}%
\providecommand \bibitemStop [0]{}%
\providecommand \bibitemNoStop [0]{.\EOS\space}%
\providecommand \EOS [0]{\spacefactor3000\relax}%
\providecommand \BibitemShut  [1]{\csname bibitem#1\endcsname}%
\let\auto@bib@innerbib\@empty
%</preamble>
\bibitem [{\citenamefont {Min}\ \emph {et~al.}(2008)\citenamefont {Min},
  \citenamefont {Zhang}, \citenamefont {Francis},\ and\ \citenamefont
  {Agnew}}]{Min2008}%
  \BibitemOpen
  \bibfield  {author} {\bibinfo {author} {\bibfnamefont {S.}~\bibnamefont
  {Min}}, \bibinfo {author} {\bibfnamefont {X.}~\bibnamefont {Zhang}}, \bibinfo
  {author} {\bibfnamefont {F.~W.}\ \bibnamefont {Francis}}, \ and\ \bibinfo
  {author} {\bibfnamefont {T.}~\bibnamefont {Agnew}},\ }\bibfield  {title}
  {\enquote {\bibinfo {title} {Human influence on arctic sea ice detectable
  from early 1990s onwards},}\ }\href@noop {} {\bibfield  {journal} {\bibinfo
  {journal} {Geophysical Research Letters}\ }\textbf {\bibinfo {volume} {35}}
  (\bibinfo {year} {2008})}\BibitemShut {NoStop}%
\bibitem [{\citenamefont {Notz}\ and\ \citenamefont
  {Marotzke}(2012)}]{Notz2012}%
  \BibitemOpen
  \bibfield  {author} {\bibinfo {author} {\bibfnamefont {D.}~\bibnamefont
  {Notz}}\ and\ \bibinfo {author} {\bibfnamefont {J.}~\bibnamefont
  {Marotzke}},\ }\bibfield  {title} {\enquote {\bibinfo {title} {Observations
  reveal external driver for arctic sea-ice retreat},}\ }\href@noop {}
  {\bibfield  {journal} {\bibinfo  {journal} {Geophysical Research Letters}\
  }\textbf {\bibinfo {volume} {39}} (\bibinfo {year} {2012})}\BibitemShut
  {NoStop}%
\bibitem [{\citenamefont {Scheweiger}, \citenamefont {Wood},\ and\
  \citenamefont {Zhang}(2019)}]{Schweiger2019}%
  \BibitemOpen
  \bibfield  {author} {\bibinfo {author} {\bibfnamefont {A.~J.}\ \bibnamefont
  {Scheweiger}}, \bibinfo {author} {\bibfnamefont {K.~R.}\ \bibnamefont
  {Wood}}, \ and\ \bibinfo {author} {\bibfnamefont {J.}~\bibnamefont {Zhang}},\
  }\bibfield  {title} {\enquote {\bibinfo {title} {Arctic sea ice volume
  variability over 1901–2010: A model-based reconstruction.}}\ }\href@noop {}
  {\bibfield  {journal} {\bibinfo  {journal} {Journal of Climate}\ }\textbf
  {\bibinfo {volume} {32}},\ \bibinfo {pages} {4731--4752} (\bibinfo {year}
  {2019})}\BibitemShut {NoStop}%
\bibitem [{\citenamefont {Budyko}(1969)}]{Budyko1969}%
  \BibitemOpen
  \bibfield  {author} {\bibinfo {author} {\bibfnamefont {M.~I.}\ \bibnamefont
  {Budyko}},\ }\bibfield  {title} {\enquote {\bibinfo {title} {The effect of
  solar radiation variations on the climate of the earth},}\ }\href@noop {}
  {\bibfield  {journal} {\bibinfo  {journal} {Tellus}\ }\textbf {\bibinfo
  {volume} {21}},\ \bibinfo {pages} {611--619} (\bibinfo {year}
  {1969})}\BibitemShut {NoStop}%
\bibitem [{\citenamefont {Sellers}(1969)}]{Sellers1969}%
  \BibitemOpen
  \bibfield  {author} {\bibinfo {author} {\bibfnamefont {W.~D.}\ \bibnamefont
  {Sellers}},\ }\bibfield  {title} {\enquote {\bibinfo {title} {A global
  climate model based on the energy balance of the earth-atmosphere system},}\
  }\href@noop {} {\bibfield  {journal} {\bibinfo  {journal} {Journal of Applied
  Meteorology}\ }\textbf {\bibinfo {volume} {8}},\ \bibinfo {pages} {392--400}
  (\bibinfo {year} {1969})}\BibitemShut {NoStop}%
\bibitem [{\citenamefont {North}, \citenamefont {Gahalan},\ and\ \citenamefont
  {Jr}(1981)}]{North1981}%
  \BibitemOpen
  \bibfield  {author} {\bibinfo {author} {\bibfnamefont {G.~R.}\ \bibnamefont
  {North}}, \bibinfo {author} {\bibfnamefont {R.~F.}\ \bibnamefont {Gahalan}},
  \ and\ \bibinfo {author} {\bibfnamefont {J.~A.~C.}\ \bibnamefont {Jr}},\
  }\bibfield  {title} {\enquote {\bibinfo {title} {Energy balance climate
  models},}\ }\href@noop {} {\bibfield  {journal} {\bibinfo  {journal} {Reviews
  of Geophysics}\ }\textbf {\bibinfo {volume} {19}},\ \bibinfo {pages}
  {91--121} (\bibinfo {year} {1981})}\BibitemShut {NoStop}%
\bibitem [{\citenamefont {Crowley}(2000)}]{Crowley2000}%
  \BibitemOpen
  \bibfield  {author} {\bibinfo {author} {\bibfnamefont {T.~J.}\ \bibnamefont
  {Crowley}},\ }\bibfield  {title} {\enquote {\bibinfo {title} {Causes of
  climate change over the past 1000 years},}\ }\href@noop {} {\bibfield
  {journal} {\bibinfo  {journal} {Science}\ }\textbf {\bibinfo {volume}
  {289}},\ \bibinfo {pages} {270--277} (\bibinfo {year} {2000})}\BibitemShut
  {NoStop}%
\bibitem [{\citenamefont {Zhang}, \citenamefont {Rothrock},\ and\ \citenamefont
  {Steele}(2000)}]{Zhang2000}%
  \BibitemOpen
  \bibfield  {author} {\bibinfo {author} {\bibfnamefont {J.}~\bibnamefont
  {Zhang}}, \bibinfo {author} {\bibfnamefont {D.}~\bibnamefont {Rothrock}}, \
  and\ \bibinfo {author} {\bibfnamefont {M.}~\bibnamefont {Steele}},\
  }\bibfield  {title} {\enquote {\bibinfo {title} {Recent changes in arctic sea
  ice: The interplay between ice dynamics and thermodynamics},}\ }\href@noop {}
  {\bibfield  {journal} {\bibinfo  {journal} {Journal of Climate}\ }\textbf
  {\bibinfo {volume} {13}},\ \bibinfo {pages} {3099--3114} (\bibinfo {year}
  {2000})}\BibitemShut {NoStop}%
\bibitem [{\citenamefont {Moon}\ and\ \citenamefont
  {Wettlaufer}(2017)}]{Moon2017}%
  \BibitemOpen
  \bibfield  {author} {\bibinfo {author} {\bibfnamefont {W.}~\bibnamefont
  {Moon}}\ and\ \bibinfo {author} {\bibfnamefont {J.~S.}\ \bibnamefont
  {Wettlaufer}},\ }\bibfield  {title} {\enquote {\bibinfo {title} {A stochastic
  dynamical model of arctic sea ice},}\ }\href@noop {} {\bibfield  {journal}
  {\bibinfo  {journal} {Journal of Climate}\ }\textbf {\bibinfo {volume}
  {30}},\ \bibinfo {pages} {5119--5140} (\bibinfo {year} {2017})}\BibitemShut
  {NoStop}%
\bibitem [{\citenamefont {Eisenman}\ and\ \citenamefont
  {Wettlaufer}(2009)}]{Eisenman2009}%
  \BibitemOpen
  \bibfield  {author} {\bibinfo {author} {\bibfnamefont {I.}~\bibnamefont
  {Eisenman}}\ and\ \bibinfo {author} {\bibfnamefont {J.~S.}\ \bibnamefont
  {Wettlaufer}},\ }\bibfield  {title} {\enquote {\bibinfo {title} {Nonlinear
  threshold behavior during the loss of arctic sea ice},}\ }\href@noop {}
  {\bibfield  {journal} {\bibinfo  {journal} {PNAS}\ }\textbf {\bibinfo
  {volume} {106}},\ \bibinfo {pages} {28--32} (\bibinfo {year}
  {2009})}\BibitemShut {NoStop}%
\bibitem [{\citenamefont {Abbot}, \citenamefont {Silber},\ and\ \citenamefont
  {Pierrehumbert}(2011)}]{Abbot2011}%
  \BibitemOpen
  \bibfield  {author} {\bibinfo {author} {\bibfnamefont {D.~S.}\ \bibnamefont
  {Abbot}}, \bibinfo {author} {\bibfnamefont {M.}~\bibnamefont {Silber}}, \
  and\ \bibinfo {author} {\bibfnamefont {R.~T.}\ \bibnamefont
  {Pierrehumbert}},\ }\bibfield  {title} {\enquote {\bibinfo {title}
  {Bifurcations leading to summer arctic sea ice loss},}\ }\href@noop {}
  {\bibfield  {journal} {\bibinfo  {journal} {Journal of Geophysical
  Research-Atmospheres}\ }\textbf {\bibinfo {volume} {116}} (\bibinfo {year}
  {2011})}\BibitemShut {NoStop}%
\bibitem [{\citenamefont {Hill}, \citenamefont {Abbot},\ and\ \citenamefont
  {Silber}(2016)}]{Hill2016}%
  \BibitemOpen
  \bibfield  {author} {\bibinfo {author} {\bibfnamefont {K.}~\bibnamefont
  {Hill}}, \bibinfo {author} {\bibfnamefont {D.~S.}\ \bibnamefont {Abbot}}, \
  and\ \bibinfo {author} {\bibfnamefont {M.}~\bibnamefont {Silber}},\
  }\bibfield  {title} {\enquote {\bibinfo {title} {Analysis of an arctic sea
  ice loss model in the limit of a discontinuous albedo},}\ }\href@noop {}
  {\bibfield  {journal} {\bibinfo  {journal} {SIAM Journal on Applied Dynamical
  Systems}\ }\textbf {\bibinfo {volume} {15}},\ \bibinfo {pages} {1163--1192}
  (\bibinfo {year} {2016})}\BibitemShut {NoStop}%
\bibitem [{\citenamefont {Gao}\ \emph {et~al.}(2016)\citenamefont {Gao},
  \citenamefont {Duan}, \citenamefont {Kan},\ and\ \citenamefont
  {Cheng}}]{Gao2016}%
  \BibitemOpen
  \bibfield  {author} {\bibinfo {author} {\bibfnamefont {T.}~\bibnamefont
  {Gao}}, \bibinfo {author} {\bibfnamefont {J.}~\bibnamefont {Duan}}, \bibinfo
  {author} {\bibfnamefont {X.}~\bibnamefont {Kan}}, \ and\ \bibinfo {author}
  {\bibfnamefont {Z.}~\bibnamefont {Cheng}},\ }\bibfield  {title} {\enquote
  {\bibinfo {title} {Dynamical inference for transitions in stochastic systems
  with $\alpha$-stable l{e}vy noise},}\ }\href@noop {} {\bibfield  {journal}
  {\bibinfo  {journal} {Journal of Physics A: Mathematical and Theoretical}\
  }\textbf {\bibinfo {volume} {49}},\ \bibinfo {pages} {294002} (\bibinfo
  {year} {2016})}\BibitemShut {NoStop}%
\bibitem [{\citenamefont {Kampen}(1976)}]{Van1976}%
  \BibitemOpen
  \bibfield  {author} {\bibinfo {author} {\bibfnamefont {N.~G.~V.}\
  \bibnamefont {Kampen}},\ }\bibfield  {title} {\enquote {\bibinfo {title}
  {Stochastic differential equation},}\ }\href@noop {} {\bibfield  {journal}
  {\bibinfo  {journal} {Physics Reports}\ }\textbf {\bibinfo {volume} {24}},\
  \bibinfo {pages} {171--228} (\bibinfo {year} {1976})}\BibitemShut {NoStop}%
\bibitem [{\citenamefont {Hasselman}(1976)}]{Hasselman1976}%
  \BibitemOpen
  \bibfield  {author} {\bibinfo {author} {\bibfnamefont {K.}~\bibnamefont
  {Hasselman}},\ }\bibfield  {title} {\enquote {\bibinfo {title} {Stochastic
  climate models. {Part I. Theory}},}\ }\href@noop {} {\bibfield  {journal}
  {\bibinfo  {journal} {Tellus}\ }\textbf {\bibinfo {volume} {28}},\ \bibinfo
  {pages} {473--485} (\bibinfo {year} {1976})}\BibitemShut {NoStop}%
\bibitem [{\citenamefont {Agarwal}\ and\ \citenamefont
  {Wettlaufer}(2018)}]{Agarwal2018}%
  \BibitemOpen
  \bibfield  {author} {\bibinfo {author} {\bibfnamefont {S.}~\bibnamefont
  {Agarwal}}\ and\ \bibinfo {author} {\bibfnamefont {J.~S.}\ \bibnamefont
  {Wettlaufer}},\ }\bibfield  {title} {\enquote {\bibinfo {title} {Fluctuations
  in arctic sea ice extent: comparing observations and climate models},}\
  }\href@noop {} {\bibfield  {journal} {\bibinfo  {journal} {Philosophical
  Transactions of Royal Society A: Mathematical, Physical and Engneering
  Sciences}\ }\textbf {\bibinfo {volume} {376}} (\bibinfo {year}
  {2018})}\BibitemShut {NoStop}%
\bibitem [{\citenamefont {Chen}\ \emph {et~al.}(2019)\citenamefont {Chen},
  \citenamefont {Gemmer}, \citenamefont {Silber},\ and\ \citenamefont
  {Volkening}}]{Chen2019}%
  \BibitemOpen
  \bibfield  {author} {\bibinfo {author} {\bibfnamefont {Y.}~\bibnamefont
  {Chen}}, \bibinfo {author} {\bibfnamefont {J.~A.}\ \bibnamefont {Gemmer}},
  \bibinfo {author} {\bibfnamefont {M.}~\bibnamefont {Silber}}, \ and\ \bibinfo
  {author} {\bibfnamefont {A.}~\bibnamefont {Volkening}},\ }\bibfield  {title}
  {\enquote {\bibinfo {title} {Noise-induced tipping under periodic forcing:
  Preferred tipping phase in a non-adiabatic forcing regime},}\ }\href@noop {}
  {\bibfield  {journal} {\bibinfo  {journal} {Chaos}\ }\textbf {\bibinfo
  {volume} {29}},\ \bibinfo {pages} {043119} (\bibinfo {year}
  {2019})}\BibitemShut {NoStop}%
\bibitem [{\citenamefont {Farazmand}\ and\ \citenamefont
  {Sapsis}(2017)}]{Farazmand2017}%
  \BibitemOpen
  \bibfield  {author} {\bibinfo {author} {\bibfnamefont {M.}~\bibnamefont
  {Farazmand}}\ and\ \bibinfo {author} {\bibfnamefont {T.~P.}\ \bibnamefont
  {Sapsis}},\ }\bibfield  {title} {\enquote {\bibinfo {title} {A variational
  approach to probing extreme events in turbulent dynamical systems},}\
  }\href@noop {} {\bibfield  {journal} {\bibinfo  {journal} {Science Advances}\
  }\textbf {\bibinfo {volume} {3}},\ \bibinfo {pages} {e1701533} (\bibinfo
  {year} {2017})}\BibitemShut {NoStop}%
\bibitem [{\citenamefont {Selmi}\ \emph {et~al.}(2016)\citenamefont {Selmi},
  \citenamefont {Coulibaly}, \citenamefont {Loghmari}, \citenamefont {Sagnes},
  \citenamefont {Beaudoin}, \citenamefont {Clerc},\ and\ \citenamefont
  {Barbay}}]{Selmi2016}%
  \BibitemOpen
  \bibfield  {author} {\bibinfo {author} {\bibfnamefont {F.}~\bibnamefont
  {Selmi}}, \bibinfo {author} {\bibfnamefont {S.}~\bibnamefont {Coulibaly}},
  \bibinfo {author} {\bibfnamefont {Z.}~\bibnamefont {Loghmari}}, \bibinfo
  {author} {\bibfnamefont {I.}~\bibnamefont {Sagnes}}, \bibinfo {author}
  {\bibfnamefont {G.}~\bibnamefont {Beaudoin}}, \bibinfo {author}
  {\bibfnamefont {M.~G.}\ \bibnamefont {Clerc}}, \ and\ \bibinfo {author}
  {\bibfnamefont {S.}~\bibnamefont {Barbay}},\ }\bibfield  {title} {\enquote
  {\bibinfo {title} {Spatiotemporal chaos induces extreme events in an extended
  microcavity laser},}\ }\href@noop {} {\bibfield  {journal} {\bibinfo
  {journal} {Physical Review Letters}\ }\textbf {\bibinfo {volume} {116}},\
  \bibinfo {pages} {013901} (\bibinfo {year} {2016})}\BibitemShut {NoStop}%
\bibitem [{\citenamefont {Ditlevsen}(1999)}]{Ditlevsen1999}%
  \BibitemOpen
  \bibfield  {author} {\bibinfo {author} {\bibfnamefont {P.~D.}\ \bibnamefont
  {Ditlevsen}},\ }\bibfield  {title} {\enquote {\bibinfo {title} {Observation
  of $\alpha$-stable noise induced millennial climate changes from an ice-core
  record},}\ }\href@noop {} {\bibfield  {journal} {\bibinfo  {journal}
  {Geophysical Research Letters}\ }\textbf {\bibinfo {volume} {26}},\ \bibinfo
  {pages} {1441--1444} (\bibinfo {year} {1999})}\BibitemShut {NoStop}%
\bibitem [{\citenamefont {Zheng}\ \emph {et~al.}(2020)\citenamefont {Zheng},
  \citenamefont {Yang}, \citenamefont {Duan}, \citenamefont {Sun},
  \citenamefont {Fu},\ and\ \citenamefont {Kurths}}]{Zheng2020}%
  \BibitemOpen
  \bibfield  {author} {\bibinfo {author} {\bibfnamefont {Y.}~\bibnamefont
  {Zheng}}, \bibinfo {author} {\bibfnamefont {F.}~\bibnamefont {Yang}},
  \bibinfo {author} {\bibfnamefont {J.}~\bibnamefont {Duan}}, \bibinfo {author}
  {\bibfnamefont {X.}~\bibnamefont {Sun}}, \bibinfo {author} {\bibfnamefont
  {L.}~\bibnamefont {Fu}}, \ and\ \bibinfo {author} {\bibfnamefont
  {J.}~\bibnamefont {Kurths}},\ }\bibfield  {title} {\enquote {\bibinfo {title}
  {The maximum likelihood climate change for global warming under the influence
  of greenhouse effect and l\'evy noise},}\ }\href@noop {} {\bibfield
  {journal} {\bibinfo  {journal} {Chaos}\ }\textbf {\bibinfo {volume} {30}},\
  \bibinfo {pages} {013132} (\bibinfo {year} {2020})}\BibitemShut {NoStop}%
\bibitem [{\citenamefont {Livina}\ and\ \citenamefont
  {Lenton}(2013)}]{Livina2013}%
  \BibitemOpen
  \bibfield  {author} {\bibinfo {author} {\bibfnamefont {V.}~\bibnamefont
  {Livina}}\ and\ \bibinfo {author} {\bibfnamefont {T.~M.}\ \bibnamefont
  {Lenton}},\ }\bibfield  {title} {\enquote {\bibinfo {title} {A recent tipping
  point in the arctic sea-ice cover: abrupt and persistent increase in the
  seasonal cycle since 2007},}\ }\href@noop {} {\bibfield  {journal} {\bibinfo
  {journal} {Cryosphere}\ }\textbf {\bibinfo {volume} {7}},\ \bibinfo {pages}
  {275--286} (\bibinfo {year} {2013})}\BibitemShut {NoStop}%
\bibitem [{\citenamefont {Lenton}\ \emph {et~al.}(2008)\citenamefont {Lenton},
  \citenamefont {Held}, \citenamefont {Kriegle}, \citenamefont {Hall},
  \citenamefont {Lucht}, \citenamefont {Rahmstor},\ and\ \citenamefont
  {Schellnhuber}}]{Leton2008}%
  \BibitemOpen
  \bibfield  {author} {\bibinfo {author} {\bibfnamefont {T.~M.}\ \bibnamefont
  {Lenton}}, \bibinfo {author} {\bibfnamefont {H.}~\bibnamefont {Held}},
  \bibinfo {author} {\bibfnamefont {E.}~\bibnamefont {Kriegle}}, \bibinfo
  {author} {\bibfnamefont {J.~W.}\ \bibnamefont {Hall}}, \bibinfo {author}
  {\bibfnamefont {W.}~\bibnamefont {Lucht}}, \bibinfo {author} {\bibfnamefont
  {S.}~\bibnamefont {Rahmstor}}, \ and\ \bibinfo {author} {\bibfnamefont
  {H.~J.}\ \bibnamefont {Schellnhuber}},\ }\bibfield  {title} {\enquote
  {\bibinfo {title} {Tipping elements in the earth's climate system},}\
  }\href@noop {} {\bibfield  {journal} {\bibinfo  {journal} {PNAS}\ }\textbf
  {\bibinfo {volume} {105}},\ \bibinfo {pages} {1786--1793} (\bibinfo {year}
  {2008})}\BibitemShut {NoStop}%
\bibitem [{\citenamefont {Ashwin}\ \emph {et~al.}(2012)\citenamefont {Ashwin},
  \citenamefont {Wieczorek}, \citenamefont {Vitolo},\ and\ \citenamefont
  {Cox}}]{Ashwin2012}%
  \BibitemOpen
  \bibfield  {author} {\bibinfo {author} {\bibfnamefont {P.}~\bibnamefont
  {Ashwin}}, \bibinfo {author} {\bibfnamefont {S.}~\bibnamefont {Wieczorek}},
  \bibinfo {author} {\bibfnamefont {R.}~\bibnamefont {Vitolo}}, \ and\ \bibinfo
  {author} {\bibfnamefont {P.}~\bibnamefont {Cox}},\ }\bibfield  {title}
  {\enquote {\bibinfo {title} {Tipping points in open systems: bifurcation,
  noise-induced and rate-independent examples in the climate system.}}\
  }\href@noop {} {\bibfield  {journal} {\bibinfo  {journal} {Philosophical
  Transactions of the Royal Society A: Mathemathical,Physical and Engineering
  Sciences}\ }\textbf {\bibinfo {volume} {370}},\ \bibinfo {pages} {1166--1184}
  (\bibinfo {year} {2012})}\BibitemShut {NoStop}%
\bibitem [{\citenamefont {Sutera}(1981)}]{Sutera1981}%
  \BibitemOpen
  \bibfield  {author} {\bibinfo {author} {\bibfnamefont {A.}~\bibnamefont
  {Sutera}},\ }\bibfield  {title} {\enquote {\bibinfo {title} {On stochastic
  perturbation and long-term climate behaviour},}\ }\href@noop {} {\bibfield
  {journal} {\bibinfo  {journal} {Quarterly Journal of the Royal Meteorological
  Society}\ }\textbf {\bibinfo {volume} {107}},\ \bibinfo {pages} {137--151}
  (\bibinfo {year} {1981})}\BibitemShut {NoStop}%
\bibitem [{\citenamefont {Lucarini}\ and\ \citenamefont
  {B\'odai}(2017)}]{Lucarini2017}%
  \BibitemOpen
  \bibfield  {author} {\bibinfo {author} {\bibfnamefont {V.}~\bibnamefont
  {Lucarini}}\ and\ \bibinfo {author} {\bibfnamefont {T.}~\bibnamefont
  {B\'odai}},\ }\bibfield  {title} {\enquote {\bibinfo {title} {Edge states in
  the climate system: exploring global instabilities and critical
  transitions},}\ }\href@noop {} {\bibfield  {journal} {\bibinfo  {journal}
  {Nonlinearity}\ }\textbf {\bibinfo {volume} {30}},\ \bibinfo {pages}
  {R32--R66} (\bibinfo {year} {2017})}\BibitemShut {NoStop}%
\bibitem [{\citenamefont {Lucarini}\ and\ \citenamefont
  {B\'odai}(2019)}]{Lucarini2019}%
  \BibitemOpen
  \bibfield  {author} {\bibinfo {author} {\bibfnamefont {V.}~\bibnamefont
  {Lucarini}}\ and\ \bibinfo {author} {\bibfnamefont {T.}~\bibnamefont
  {B\'odai}},\ }\bibfield  {title} {\enquote {\bibinfo {title} {Transitions
  accross melancholia states in a climate model: Reconciling the deterministic
  and stochastic points of view},}\ }\href@noop {} {\bibfield  {journal}
  {\bibinfo  {journal} {Physical Review Letters}\ }\textbf {\bibinfo {volume}
  {122}},\ \bibinfo {pages} {158701} (\bibinfo {year} {2019})}\BibitemShut
  {NoStop}%
\bibitem [{\citenamefont {Duan}(2015)}]{Duan2015}%
  \BibitemOpen
  \bibfield  {author} {\bibinfo {author} {\bibfnamefont {J.}~\bibnamefont
  {Duan}},\ }\href@noop {} {\emph {\bibinfo {title} {An Introduction to
  Stochastic Dynamics}}}\ (\bibinfo  {publisher} {Cambridge University Press},\
  \bibinfo {year} {2015})\BibitemShut {NoStop}%
\bibitem [{\citenamefont {Sun}\ and\ \citenamefont {Duan}(2012)}]{Sun2012}%
  \BibitemOpen
  \bibfield  {author} {\bibinfo {author} {\bibfnamefont {X.}~\bibnamefont
  {Sun}}\ and\ \bibinfo {author} {\bibfnamefont {J.}~\bibnamefont {Duan}},\
  }\bibfield  {title} {\enquote {\bibinfo {title} {{Fokker-Planck} equations
  for nonlinear dynamical systems driven by {non-Gaussian L\'evy} processes},}\
  }\href@noop {} {\bibfield  {journal} {\bibinfo  {journal} {Journal of
  Mathematical Physics}\ }\textbf {\bibinfo {volume} {53}},\ \bibinfo {pages}
  {072701} (\bibinfo {year} {2012})}\BibitemShut {NoStop}%
\bibitem [{\citenamefont {Gao}, \citenamefont {Duan},\ and\ \citenamefont
  {Li}(2016)}]{Gao2016b}%
  \BibitemOpen
  \bibfield  {author} {\bibinfo {author} {\bibfnamefont {T.}~\bibnamefont
  {Gao}}, \bibinfo {author} {\bibfnamefont {J.}~\bibnamefont {Duan}}, \ and\
  \bibinfo {author} {\bibfnamefont {X.}~\bibnamefont {Li}},\ }\bibfield
  {title} {\enquote {\bibinfo {title} {{Fokker-Planck} equations for stochastic
  dynamical systems with symmetric {L\'evy} motions},}\ }\href@noop {}
  {\bibfield  {journal} {\bibinfo  {journal} {Applied Mathematics and
  Computation}\ }\textbf {\bibinfo {volume} {278}},\ \bibinfo {pages} {1--20}
  (\bibinfo {year} {2016})}\BibitemShut {NoStop}%
\bibitem [{\citenamefont {Zeitouni}\ and\ \citenamefont
  {Dembo}(1987)}]{Zeitouni1987}%
  \BibitemOpen
  \bibfield  {author} {\bibinfo {author} {\bibfnamefont {O.}~\bibnamefont
  {Zeitouni}}\ and\ \bibinfo {author} {\bibfnamefont {A.}~\bibnamefont
  {Dembo}},\ }\bibfield  {title} {\enquote {\bibinfo {title} {A maximum a
  posteriori estimator for trajectories of diffusion processes},}\ }\href@noop
  {} {\bibfield  {journal} {\bibinfo  {journal} {Stochastics}\ }\textbf
  {\bibinfo {volume} {20}},\ \bibinfo {pages} {221--246} (\bibinfo {year}
  {1987})}\BibitemShut {NoStop}%
\bibitem [{\citenamefont {Zeitouni}\ and\ \citenamefont
  {Dembo}(1988)}]{Zeitouni1988}%
  \BibitemOpen
  \bibfield  {author} {\bibinfo {author} {\bibfnamefont {O.}~\bibnamefont
  {Zeitouni}}\ and\ \bibinfo {author} {\bibfnamefont {A.}~\bibnamefont
  {Dembo}},\ }\bibfield  {title} {\enquote {\bibinfo {title} {An existence
  theorem and some properties of maximum a posteriori estimators of
  trajectories of diffusions},}\ }\href@noop {} {\bibfield  {journal} {\bibinfo
   {journal} {Stochastics}\ }\textbf {\bibinfo {volume} {23}},\ \bibinfo
  {pages} {197--218} (\bibinfo {year} {1988})}\BibitemShut {NoStop}%
\bibitem [{\citenamefont {Maykut}\ and\ \citenamefont
  {Untersteiner}(1971)}]{Maykut1971}%
  \BibitemOpen
  \bibfield  {author} {\bibinfo {author} {\bibfnamefont {G.~A.}\ \bibnamefont
  {Maykut}}\ and\ \bibinfo {author} {\bibfnamefont {N.}~\bibnamefont
  {Untersteiner}},\ }\bibfield  {title} {\enquote {\bibinfo {title} {Some
  results from a time-dependent thermodynamic model of sea ice},}\ }\href@noop
  {} {\bibfield  {journal} {\bibinfo  {journal} {Journal of Geophysical
  Research}\ }\textbf {\bibinfo {volume} {76}},\ \bibinfo {pages} {1550--1575}
  (\bibinfo {year} {1971})}\BibitemShut {NoStop}%
\bibitem [{\citenamefont {Fetterer}\ and\ \citenamefont
  {Untersteiner}(1998)}]{Fetterer1998}%
  \BibitemOpen
  \bibfield  {author} {\bibinfo {author} {\bibfnamefont {F.}~\bibnamefont
  {Fetterer}}\ and\ \bibinfo {author} {\bibfnamefont {N.}~\bibnamefont
  {Untersteiner}},\ }\bibfield  {title} {\enquote {\bibinfo {title}
  {Observations of melt ponds on arctic sea ice},}\ }\href@noop {} {\bibfield
  {journal} {\bibinfo  {journal} {Journal of Geophysical Research-Oceans}\
  }\textbf {\bibinfo {volume} {103}},\ \bibinfo {pages} {24821--24835}
  (\bibinfo {year} {1998})}\BibitemShut {NoStop}%
\bibitem [{\citenamefont {Kerr}(2012)}]{Kerr2012}%
  \BibitemOpen
  \bibfield  {author} {\bibinfo {author} {\bibfnamefont {R.~A.}\ \bibnamefont
  {Kerr}},\ }\bibfield  {title} {\enquote {\bibinfo {title} {Ice-free arctic
  sea may be years, not decades, away},}\ }\href@noop {} {\bibfield  {journal}
  {\bibinfo  {journal} {Science}\ }\textbf {\bibinfo {volume} {337}},\ \bibinfo
  {pages} {1591--1591} (\bibinfo {year} {2012})}\BibitemShut {NoStop}%
\bibitem [{\citenamefont {Stroeve}\ \emph {et~al.}(2005)\citenamefont
  {Stroeve}, \citenamefont {Serreze}, \citenamefont {Fetterer}, \citenamefont
  {Arbetter}, \citenamefont {Meier}, \citenamefont {Maslanik},\ and\
  \citenamefont {Knowles}}]{Stroeve2005}%
  \BibitemOpen
  \bibfield  {author} {\bibinfo {author} {\bibfnamefont {J.~C.}\ \bibnamefont
  {Stroeve}}, \bibinfo {author} {\bibfnamefont {M.~C.}\ \bibnamefont
  {Serreze}}, \bibinfo {author} {\bibfnamefont {F.}~\bibnamefont {Fetterer}},
  \bibinfo {author} {\bibfnamefont {T.}~\bibnamefont {Arbetter}}, \bibinfo
  {author} {\bibfnamefont {W.}~\bibnamefont {Meier}}, \bibinfo {author}
  {\bibfnamefont {J.}~\bibnamefont {Maslanik}}, \ and\ \bibinfo {author}
  {\bibfnamefont {K.}~\bibnamefont {Knowles}},\ }\bibfield  {title} {\enquote
  {\bibinfo {title} {Tracking the arctic's shriking ice cover: Another extreme
  september minimum in 2004},}\ }\href@noop {} {\bibfield  {journal} {\bibinfo
  {journal} {Geophysical Research Letters}\ }\textbf {\bibinfo {volume} {32}}
  (\bibinfo {year} {2005})}\BibitemShut {NoStop}%
\bibitem [{\citenamefont {Walker}(2016)}]{Walker2016}%
  \BibitemOpen
  \bibfield  {author} {\bibinfo {author} {\bibfnamefont {G.}~\bibnamefont
  {Walker}},\ }\bibfield  {title} {\enquote {\bibinfo {title} {The tipping
  point of the iceberg},}\ }\href@noop {} {\bibfield  {journal} {\bibinfo
  {journal} {Nature}\ }\textbf {\bibinfo {volume} {441}},\ \bibinfo {pages}
  {802--805} (\bibinfo {year} {2016})}\BibitemShut {NoStop}%
\bibitem [{\citenamefont {Bitz}\ and\ \citenamefont {Roe}(2004)}]{Bitz2004}%
  \BibitemOpen
  \bibfield  {author} {\bibinfo {author} {\bibfnamefont {C.~M.}\ \bibnamefont
  {Bitz}}\ and\ \bibinfo {author} {\bibfnamefont {G.~H.}\ \bibnamefont {Roe}},\
  }\bibfield  {title} {\enquote {\bibinfo {title} {A mechanism for the high
  rate of sea ice thinning in the arctic ocean},}\ }\href@noop {} {\bibfield
  {journal} {\bibinfo  {journal} {Journal of Climate}\ }\textbf {\bibinfo
  {volume} {17}},\ \bibinfo {pages} {3623--3632} (\bibinfo {year}
  {2004})}\BibitemShut {NoStop}%
\bibitem [{\citenamefont {Muller-Stoffels}\ and\ \citenamefont
  {Wackerbauer}(2011)}]{Mueller2011}%
  \BibitemOpen
  \bibfield  {author} {\bibinfo {author} {\bibfnamefont {M.}~\bibnamefont
  {Muller-Stoffels}}\ and\ \bibinfo {author} {\bibfnamefont {R.}~\bibnamefont
  {Wackerbauer}},\ }\bibfield  {title} {\enquote {\bibinfo {title} {Regular
  network model for the sea ice-albedo feedback in the arctic},}\ }\href@noop
  {} {\bibfield  {journal} {\bibinfo  {journal} {Chaos}\ }\textbf {\bibinfo
  {volume} {21}} (\bibinfo {year} {2011})}\BibitemShut {NoStop}%
\bibitem [{\citenamefont {Saltzman}(2002)}]{Saltzman2002}%
  \BibitemOpen
  \bibfield  {author} {\bibinfo {author} {\bibfnamefont {B.}~\bibnamefont
  {Saltzman}},\ }\href@noop {} {\emph {\bibinfo {title} {Dynamical
  Paleoclimatology}}}\ (\bibinfo  {publisher} {Academic Press},\ \bibinfo
  {year} {2002})\BibitemShut {NoStop}%
\bibitem [{\citenamefont {Yang}(2020)}]{Yang2020}%
  \BibitemOpen
  \bibfield  {author} {\bibinfo {author} {\bibfnamefont {F.}~\bibnamefont
  {Yang}},\ }\href@noop {} {\enquote {\bibinfo {title} {Code},}\ }\bibinfo
  {howpublished} {Github} (\bibinfo {year} {2020}),\ \bibinfo {note}
  {\url{https://github.com/yangfang0914/The-tipping-times-in-an-arctic-sea-ice-system-under-influence-of-extreme-events}}\BibitemShut
  {NoStop}%
\end{thebibliography}


\end{document}